\documentclass[npg,ms]{copernicus}

\usepackage{natbib}

\usepackage{epsfig}

\usepackage{amssymb,latexsym,times, float}
\usepackage[hang,small]{caption}

\usepackage{algorithmic, algorithm}

\begin{document}
\title{Monte Carlo fixed-lag smoothing in state-space models}
\author[1,2]{Anne Cuzol}
\author[1]{Etienne M\'emin}
\affil[1]{INRIA Rennes-Bretagne Atlantique, Rennes, France}
\affil[2]{LMAM, Laboratory of Mathematics and Applications of Mathematics, University of Bretagne Sud, Vannes, France}

\maketitle

\begin{abstract}
 
This paper presents an algorithm for Monte Carlo fixed-lag smoothing in state-space models defined by a  diffusion process observed through noisy discrete-time measurements. Based on a particles approximation of the filtering and smoothing distributions, the method relies on a simulation technique of conditioned diffusions. The proposed sequential smoother can be applied to general non linear and multidimensional models, like the ones used in environmental applications. The smoothing of a turbulent flow in a high-dimensional context is given as a practical example.\\


\end{abstract}

\introduction

The framework of this paper concerns state-space models described by general diffusions of the form:
\begin{equation}
{\rm d}\mathbf{x}(t)=f(\mathbf{x}(t)){\rm d}t + \sigma(\mathbf{x}(t)) {\rm d}\mathbf{B}(t),
\label{Diffusion_model}
\end{equation}
which are partially observed through noisy measurements at discrete times. Such models can describe many dynamical phenomena in environmental sciences, physics, but also in finance or engineering applications. The main motivation of this work concerns environmental  applications, where  non linearity and high-dimensionality arise. Indeed,  environmental models and data describe  non linear phenomena over large domains, with  high spatial resolution. The continuous dynamical model (\ref{Diffusion_model}) is defined from \textit{a priori} physical laws, while observations are supplied by sensors (satellite data for instance) and can appear with very low time frequency. As an example, in the application presented in the last part of this paper, the dimension of the state and observations is of the order of many thousands, and the model is described by the non linear Navier-Stokes equation. Filtering and smoothing in such state space models aim at coupling model and observations, which is called data assimilation. The goal of the filtering is to estimate the system state distribution knowing past and present observations. This allows for instance to give proper initial conditions to forecast the future state of a system characterizing atmospheric or oceanographic flows. On the other hand, the smoothing aims at estimating the state distribution using past and future observations, and this retrospective state estimation allows to analyze a spatio-temporal phenomenon over a given time period, for climatology studies for instance. 
Applications of data assimilation are numerous and the interest is growing in environmental sciences with the increase of available data. However, it is still a challenge to develop filtering and smoothing methods that can be used within a general non linear and high-dimensional context.

Monte Carlo sequential methods, contrary to standard Kalman filters,  are able to deal with the filtering problem in non linear state-space models. The particle filtering  \citep{DelMoral01,Doucet00} solves the whole  filtering equations through Monte Carlo approximations of the state distribution. On the other hand, ensemble Kalman methods \citep{Evensen03}  take into account in some way the non linearities in  the system, but are  based on a Gaussian assumption. For high-dimensional systems, ensemble Kalman methods are preferred in practice to particle filters \citep{Stroud10,VanLeeuwen09} since they reach better performance for limited number of particles. In order to keep this advantage while alleviating the Gaussian assumption, both methods are combined in \citet{Papadakis10}, leading to a particle filter that can  be applied to high-dimensional systems. We will use this  technique for the filtering step in the  high-dimensional application presented in Section \ref{Application}.

The aim of this paper is to propose a new smoothing method. It is known  that within the particle filter framework, the smoothing can be computed backward, reweighting past particles using present observations \citep{Briers10,Godsill04}. There are however two main difficulties. Firstly, it is necessary to know the transition density of the process between observation times, which is not available for general diffusions. This transition density can be approximated through Monte Carlo simulations, as proposed by \citet{Durham02} to solve inference problems for diffusion processes. However, these approximations are based on Brownian bridge (or modified versions of it) simulations, that do not take into account the drift part of the model. For non linear and  high-dimensional models with a drift term that dominates, such approximations will be inefficient. It is also possible to obtain  an unbiased estimate of the transition density (see \citet{Beskos06}), but this approach is not adapted to a  multi-dimensional context. As a matter of fact, the use of this technique in a multivariate setting imposes  constraints on the diffusion drift (in particular the drift function has to be of gradient type). Secondly, since these smoothing schemes rely on existing particles only, the estimation of smoothing distributions may become poor in a high-dimensional context, when for computational reasons the number of particles is reduced. On the other hand, in the framework of ensemble Kalman methods, \citet{Evensen00} have also proposed to estimate backward the smoothing distribution in a recursive way, based on existing filtering trajectories. \citet{Stroud10} presented and applied an ensemble Kalman smoothing method, relying on a linearization of the system dynamics.

All previously mentioned smoothing methods require to perform specific assumptions or simplifications in order to deal with general non linear models of type (\ref{Diffusion_model}) in a high-dimensional context. To the best of our knowledge, it remains  a challenging problem to develop  smoothing methods that can be used in this general setting. In this paper, we deal with this issue sequentially each time a new observation is available, by smoothing the hidden state from this new observation time up to the previous one. This approach, called fixed-lag smoothing, constitutes then a partial answer to the global smoothing problem that would take into account all available observations. Nevertheless, it is reasonable to assume that the distribution of the hidden state depends on future observations through the next observation only, as soon as the time step between measurements is long (which is typically the case in the environmental applications that motivate this work). Under this assumption, a new observation will impact the distribution of the hidden process up to the previous observation only. This point of view justifies the use of a fixed-lag smoothing in our setting as a reasonable approximation of the global smoothing problem.

For low-dimensional systems, it is known that such a fixed-lag smoothing may be directly obtained from the particle filtering result, reweighting past trajectories. However, in a high-dimensional context where, for computational reasons, the number of particles has to be reduced, this method leads to poor smoothing distribution estimates. In contrast, our method does not rely on existing particles only. It is  built on a  conditional simulation technique of diffusions proposed by \citet{Delyon06} that provides new state trajectories at hidden times between observations. This technique is adapted to a multivariate context where the drift dominates, contrary to simulation techniques based on Brownian bridge sampling \citep{Durham02}. Moreover, it does not require constraining assumptions for multivariate models, contrary to other techniques based on exact simulation of diffusions \citep{Beskos05,Beskos06}. The proposed smoothing method can then be applied to high-dimensional systems. Finally, it does not require model linearization nor Gaussian hypotheses, and so is able to deal with general non linear models.

The remaining of the paper is organized as follows. Section \ref{Filtering_smoothing} briefly presents  the filtering and smoothing problems within our state-space model framework, and focuses on the fixed-lag smoothing  that will be at the heart of this paper. Section \ref{Conditional_smoothing} presents the conditional simulation technique of diffusions of \citet{Delyon06}, and details the construction of the proposed Monte Carlo estimate of smoothing distributions. The method is then experimented on a one-dimensional example in Section \ref{Simulation_study}. Finally, the method is applied in section  \ref{Application} to a practical non linear and high-dimensional case, similar to the problems that have to be faced in environmental applications. A discussion is given in Section \ref{Discussion}.

\section{Monte Carlo  filtering and smoothing in state-space models}
\label{Filtering_smoothing}
 
In this section we present the general state-space model that defines our framework, and recall briefly the filtering and smoothing problems. In particular, we present the particle filter and the associated fixed-lag smoothing problem on which the paper concentrates.

\subsection{Framework and particle-based methods}
We are interested in continuous-discrete state-space models of the following form:
\begin{eqnarray}
{\rm d}\mathbf{x}(t)&=&f(\mathbf{x}(t)){\rm d}t + \sigma(\mathbf{x}(t)) {\rm d}\mathbf{B}(t),
\label{State_space_model_1}\\
\mathbf{y}(t_k)&=&g(\mathbf{x}(t_k))+\mathbf{\gamma}_{t_k},
\label{State_space_model_2}
\end{eqnarray}
where  the hidden state vector $\mathbf x \in \mathbb R^n$ is observed through the observation vector $\mathbf y \in \mathbb R^m$ at discrete times $\{t_1,t_2,\ldots \}$, and the drift function $f$ and observation operator $g$ can be non linear. The dynamical model uncertainty is described by a n-dimensional Brownian motion with covariance $\Sigma=\sigma(\mathbf{x}(t))\sigma(\mathbf{x}(t))^T$. The functions $f$, $g$ and $\sigma$ are assumed to be known, as well as the law of the observation noise $\mathbf{\gamma}_{t_k}$.\\

The filtering problem in such state-space models can be solved with a Monte Carlo sequential approach, called particle filtering \citep{DelMoral01,Doucet00},  allowing  the recursive estimation of the filtering distribution $p(\mathbf x_{t_1:t_k}|\mathbf y_{t_1:t_k})$, and in particular of its marginal distribution $p(\mathbf x_{t_k}|\mathbf y_{t_1:t_k})$, at each observation time $t_k$. The method relies on a Monte Carlo approximation of this distribution over a set of weighted trajectories $\{\mathbf x^{(i)}_{t_1:t_k}\}_{i=1:N}$ (called particles):
\begin{equation}
\hat{p}(\mathbf x_{t_1:t_k}|\mathbf y_{t_1:t_k})=\sum_{i=1}^{N}w_{t_k}^{(i)}\delta_{\mathbf x_{t_1:t_k}^{(i)}}(\mathbf x_{t_1:t_k}),
\label{Filtering_distribution_trajectories}
\end{equation}
whose marginal distribution at time $t_k$ writes:
\begin{equation}
\hat{p}(\mathbf x_{t_k}|\mathbf y_{t_1:t_k})=\sum_{i=1}^{N}w_{t_k}^{(i)}\delta_{\mathbf x_{t_k}^{(i)}}(\mathbf x_{t_k}).
\label{Filtering_distribution_particles}
\end{equation}

\noindent Particle filters rely on a sequential importance sampling scheme that recursively samples particles, and updates their weights at observation times. The importance sampling distribution is chosen in such a way that the importance weights $w_{t_k}^{(i)}$ can be evaluated recursively in time as observations become available, through the likelihood $p(\mathbf{y}(t_k)|\mathbf{x}(t_k))$. In practice, a resampling procedure is added in order to avoid degeneracy. This procedure duplicates trajectories with large weights and remove small weighted trajectories.

\noindent Note that the particle filtering technique updates the filtering distribution at observation times only. However, after the estimate $\hat{p}(\mathbf x_{t_k}|\mathbf y_{t_1:t_k})$ has been updated at observation time $t_k$, the filtering distribution can be predicted in order to have a continuous estimation of $\hat{p}(\mathbf x_{t}|\mathbf y_{t_1:t_k})$ for all $t \in ]t_k,t_{k+1}[$ until the next observation time:
\begin{equation}
\hat{p}(\mathbf x_{t}|\mathbf y_{t_1:t_k})=\sum_{i=1}^{N}w_{t_k}^{(i)}\delta_{\mathbf x_{t}^{(i)}}(\mathbf x_{t}),
\label{Predictive_filtering_distribution}
\end{equation}
where, for all $i=1,\ldots,N$, the state $\mathbf x_{t}^{(i)}$ is  sampled from (\ref{State_space_model_1}), starting from $\mathbf x_{t_k}^{(i)}$.\\

Contrary to the filtering approach that uses past and present observations, a  global  smoothing  in state-space models aims at estimating $p(\mathbf x_{t}|\mathbf y_{t_1:t_{\text{end}}})$ for all $t \in [t_1,t_{\text{end}}]$, using all past and future observations over a given time period. As raised in the introduction, existing smoothing methods do not apply directly to a general non linear model of type (\ref{State_space_model_1})-(\ref{State_space_model_2}) in a high-dimensional context, since  assumptions have to be made that may not be realistic. Instead of solving the global smoothing, we will  concentrate in the rest of the paper on a fixed-lag  smoothing, which constitutes a partial answer to the global smoothing problem.

\subsection{Basic particles fixed-lag smoothing}
\label{Particles_smoothing}

The objective of the fixed-lag smoothing will be to replace the predictive  distribution (\ref{Predictive_filtering_distribution}) by its smoothed version $p(\mathbf x_{t}|\mathbf y_{t_1:t_{k+1}})$ $\forall t \in ]t_k,t_{k+1}]$, sequentially each time a new observation $\mathbf y_{t_{k+1}}$ arrives. This will allow to reduce the temporal discontinuities inherent to the filtering technique, that successively  predicts the distribution of the state between observations, and updates this distribution at observation times. 

To achieve this, by construction of the particle filter that weights entire trajectories (see equation (\ref{Filtering_distribution_trajectories})), it is known (see for instance \citet{Doucet00}) that the fixed-lag smoothing distribution $\hat p(\mathbf x_{t}|\mathbf y_{t_1:t_{k+1}})$ can be directly obtained from the marginal at time $t$ of $\hat p(\mathbf x_{t_1:t_{k+1}}|\mathbf y_{t_1:t_{k+1}})$. The empirical smoothing distribution is then given by:
\begin{equation}
\hat{p}(\mathbf x_{t}|\mathbf y_{t_1:t_{k+1}})=\sum_{i=1}^{N}w_{t_{k+1}}^{(i)}\delta_{\mathbf x_{t}^{(i)}}(\mathbf x_{t}) \quad \forall t \in]t_k,t_{k+1}].
\label{Local_smoothing_particles}
\end{equation}

\noindent However, this approximation is simply a reweighting of past existing particle trajectories, so that the support of the empirical smoothing distribution is the same as the filtering one. This approximation can lead to poor estimates since it relies on the support of the filtering distribution at time $t_k$. If the number of particles is too small with respect to the state dimension, the support may be greatly reduced by the correction step (assigning small weights to all particles except a few), leading in practice to a bad estimation of $p(\mathbf x_{t}|\mathbf y_{t_1:t_{k+1}})$. Since we are interested in smoothing techniques that are efficient in a high-dimensional context,  this direct smoothing technique can not be used in its basic form and has to be improved.

In the following, we propose  to use a conditional simulation technique of diffusions that will enable the sampling of new smoothed trajectories between times $t_{k}$ and $t_{k+1}$. The  approximation of the smoothing distribution (\ref{Local_smoothing_particles}) at each hidden time will then be improved. The conditional simulation technique is presented in the next section, before the resulting smoothing procedure we propose.

\section{Fixed-lag smoothing with conditional simulation}
\label{Conditional_smoothing}

The smoothing method  we propose is based on a conditional simulation technique that is presented in section \ref{Conditional_simulation}. We develop then in section \ref{Conditional_smoothing_sub} how this technique can be used to improve the estimation of the  smoothing distribution (\ref{Local_smoothing_particles}).

\subsection{Conditional simulation}
\label{Conditional_simulation}
Conditional simulation of a diffusion aims at sampling  trajectories from a given process:
\begin{equation}
{\rm d}\mathbf{x}(t)=f(\mathbf{x}(t)){\rm d}t + \sigma(\mathbf{x}(t)) {\rm d}\mathbf{B}(t)
\label{Diffusion_process}
\end{equation}
between two times $t=0$ and $t=T$, with the constraints $\mathbf x(0)=\mathbf u$ and $\mathbf x(T)=\mathbf v$. This simulation problem is treated by \citet{Delyon06}, where the authors show how to obtain the law of the constrained process from a Girsanov theorem. In practice, the proposed algorithms consist in simulating trajectories according to another diffusion process, which is built to respect the constraints and is easy to simulate from. The conditional distribution of the constrained process (\ref{Diffusion_process}) is shown to be absolutely continuous with respect to the distribution of the auxiliary process, with explicitly given density. For instance, in the case where the drift is bounded (a similar algorithm is proposed in \citet{Delyon06} for the unbounded case) and for $\sigma$ invertible, the algorithm is based on the simulation of trajectories from the following process:
\begin{equation}
{\rm d}\mathbf{\tilde x}(t)=\left(f(\mathbf{\tilde x}(t)) -\frac{\mathbf{\tilde x}(t)-\mathbf v}{T-t}\right){\rm d}t + \sigma(\mathbf{\tilde x}(t)) {\rm d}\mathbf{B}(t),
\label{Diffusion_process_constraint}
\end{equation}
with initial condition $\mathbf{\tilde x}(0)=\mathbf u$. This process is a simple modification of (\ref{Diffusion_process}), where a deterministic part is added to the drift. It is then easy to simulate unconditional trajectories from this  process, and all simulated trajectories will satisfy $\mathbf{\tilde x}(T)=\mathbf v$ by construction. 
\noindent For simplicity we will assume in the following that $\sigma$ is independent of $\mathbf x(t)$ (note however that this is not an assumption in \citet{Delyon06}). The law of the conditioned process is given by:
\begin{equation}
\mathbb E[h(\mathbf x)|\mathbf x(0)=\mathbf u,\mathbf x(T)=\mathbf v]=\mathbb E\left[h(\mathbf{\tilde x})\alpha(\mathbf{\tilde x})\right],
\end{equation}
for all measurable function $h$,  where:
\begin{equation}
\alpha(\mathbf{\tilde x})=\exp\left(-\int_0^T\frac{(\mathbf{\tilde x}(t)-\mathbf v)^T\Sigma^{-1}f(\mathbf{\tilde x}(t))}{T-t}{\rm d}t\right)
\label{Girsanov_weights}
\end{equation}
is the density coming from  Girsanov theorem (see \citet{Delyon06}), with $\Sigma=\sigma(\mathbf{\tilde x}(t))\sigma(\mathbf{\tilde x}(t))^T$.

Let us note that the presence of the drift part of model (\ref{Diffusion_process}) in the auxiliary process (\ref{Diffusion_process_constraint}) is crucial to make the simulation efficient. The same  process had initially been proposed by \citet{Clark90} to solve the conditional simulation problem. On the other hand, standard Brownian bridges that could be used as auxiliary processes \citep{Durham02} lead in practice to poor approximations of the original constrained diffusion in our high-dimensional setting, since Brownian bridge trajectories are too far away from trajectories of (\ref{Diffusion_process}).

In the following, the conditional marginal of interest $p(\mathbf x_{t}|\mathbf x(0)=\mathbf u,\mathbf x(T)=\mathbf v)$ will then be approximated as follows:
\begin{equation}
\hat{p}(\mathbf x_{t}|\mathbf x(0)=\mathbf u,\mathbf x(T)=\mathbf v)=\sum_{j=1}^{M}\alpha(\mathbf{\tilde x}^{(j)})\delta_{\mathbf {\tilde x}_{t}^{(j)}}(\mathbf {x}_{t}) \quad \forall t \in [0,T],
\label{Conditional_distribution}
\end{equation}
where the $M$ trajectories $\{\mathbf {\tilde x}_t^{(j)}\}_{j=1:M}$ are simulated from (\ref{Diffusion_process_constraint}) with $\mathbf {\tilde x}_0^{(j)}=\mathbf u$ for all $j=1,\ldots,M$.

\subsection{Proposed fixed-lag smoothing method}
\label{Conditional_smoothing_sub}

We show in the following how the conditional simulation technique can be used to improve the estimation of the local smoothing distribution $p(\mathbf x_{t}|\mathbf y_{t_1:t_{k+1}})$ for all $t \in ]t_{k},t_{k+1}]$. 

\noindent We first note that this distribution can be decomposed as:
\begin{eqnarray}
p(\mathbf x_{t}|\mathbf y_{t_1:t_{k+1}})&=&\int p(\mathbf x_{t},\mathbf x_{t_{k}},\mathbf x_{t_{k+1}}|\mathbf y_{t_1:t_{k+1}}) {\rm d}\mathbf x_{t_{k}} {\rm d}\mathbf x_{t_{k+1}} \nonumber \\
&=&\int p(\mathbf x_{t_{k}},\mathbf x_{t_{k+1}}|\mathbf y_{t_1:t_{k+1}}) p(\mathbf x_{t}|\mathbf x_{t_{k}},\mathbf x_{t_{k+1}},\mathbf y_{t_1:t_{k+1}}) {\rm d}\mathbf x_{t_{k}}{\rm d}\mathbf x_{t_{k+1}}.
\label{eq_1}
\end{eqnarray}

\noindent Then, from the state-space model properties, we obtain:
\begin{equation}
p(\mathbf x_{t}|\mathbf y_{t_1:t_{k+1}})=\int p(\mathbf x_{t_{k}},\mathbf x_{t_{k+1}}|\mathbf y_{t_1:t_{k+1}}) p(\mathbf x_{t}|\mathbf x_{t_{k}},\mathbf x_{t_{k+1}}) {\rm d}\mathbf x_{t_{k}}{\rm d}\mathbf x_{t_{k+1}}.
\label{eq_3}
\end{equation}

\noindent Moreover, from the particle filter Monte Carlo approximation described by (\ref{Filtering_distribution_trajectories}), the joint law $p(\mathbf x_{t_{k}},\mathbf x_{t_{k+1}}|\mathbf y_{t_1:t_{k+1}})$ can be replaced by:
\begin{equation}
\hat{p}(\mathbf x_{t_{k}},\mathbf x_{t_{k+1}}|\mathbf y_{t_1:t_{k+1}})=\sum_{i=1}^N w_{t_{k+1}}^{(i)}\delta_{(\mathbf x_{t_{k+1}}^{(i)},\mathbf x_{t_{k}}^{(i)})}(\mathbf x_{t_{k+1}},\mathbf x_{t_{k}}),
\label{eq_2}
\end{equation}
where the $w_{t_{k+1}}^{(i)}$ are the particle filter importance weights.

\noindent Plugging (\ref{eq_2}) into (\ref{eq_3}) leads then to the following approximation for the fixed-lag smoothing distribution:
\begin{equation}
\hat p(\mathbf x_{t}|\mathbf y_{t_1:t_{k+1}})=\sum_{i=1}^N w_{t_{k+1}}^{(i)}p(\mathbf x_{t}|\mathbf x_{t_{k}}^{(i)},\mathbf x_{t_{k+1}}^{(i)}).
\end{equation}
The conditional distribution $p(\mathbf x_{t}|\mathbf x_{t_{k}}^{(i)},\mathbf x_{t_{k+1}}^{(i)})$ can be estimated using (\ref{Conditional_distribution}) for each pair of initial and end points $\mathbf x_{t_{k}}^{(i)}$ and $\mathbf x_{t_{k+1}}^{(i)}$:
\begin{equation}
\hat{p}(\mathbf x_{t}|\mathbf x_{t_{k}}^{(i)},\mathbf x_{t_{k+1}}^{(i)})=\sum_{j=1}^M  \alpha(\mathbf{\tilde x}^{(i)(j)}) \delta_{\mathbf{\tilde x}_t^{(i)(j)}}(\mathbf{x}_t),
\end{equation}
where each $\mathbf{\tilde x}_t^{(i)(j)}$  is sampled from (\ref{Diffusion_process_constraint}) with initial constraint $\mathbf{\tilde x}_{t_{k}}^{(i)(j)}=\mathbf x_{t_{k}}^{(i)}$ and final constraint $\mathbf x_{t_{k+1}}^{(i)}$.

\noindent The estimation of the  smoothing distribution of interest writes finally:
\begin{equation}
\hat p(\mathbf x_{t}|\mathbf y_{t_1:t_{k+1}})=\sum_{i=1}^N w_{t_{k+1}}^{(i)} \sum_{j=1}^M  \alpha(\mathbf{\tilde x}^{(i)(j)}) \delta_{\mathbf{\tilde x}_t^{(i)(j)}}(\mathbf{x}_t), \quad \forall t \in ]t_{k},t_{k+1}].
\label{Conditional_smoothing_particles}
\end{equation}

~\\
The algorithm we propose to compute the fixed-lag smoothing  distribution on a given time interval $]t_{k},t_{k+1}]$ is therefore the following:\\

\begin{algorithm}[H]
\caption{Fixed-lag conditional smoothing}
\vspace{0.2cm}
~~ For each $t_k=t_1,t_2,\ldots$:
\begin{itemize}
	\item Store $\{\mathbf x_{t_{k}}^{(i)}\}_{i=1:N}$ and compute $\{\mathbf x_{t_{k+1}}^{(i)}\}_{i=1:N}$ and associated weights $\{w_{t_{k+1}}^{(i)}\}_{i=1:N}$ from a  particle filter algorithm;
	\item For each pair  $\{\mathbf x_{t_{k}}^{(i)}, \mathbf x_{t_{k+1}}^{(i)}\}$, $i=1,\ldots,N$:
	
	\begin{itemize}
	\item Simulate $M$ conditional trajectories $\{\mathbf{\tilde x}_t^{(i)(j)}\}_{j=1:M}$ for $t \in [t_{k},t_{k+1}]$ from (\ref{Diffusion_process_constraint}) with an Euler scheme, with the constraints $\mathbf{\tilde x}_{t_{k}}^{(i)(j)}=\mathbf x_{t_{k}}^{(i)}$ and $\mathbf{\tilde x}_{t_{k+1}}^{(i)(j)}=\mathbf x_{t_{k+1}}^{(i)}$,
	\item Compute weights $\alpha(\mathbf{\tilde x}^{(i)(j)})$ from (\ref{Girsanov_weights}) for all $j=1,\ldots,M$ , with final constraint $\mathbf x_{t_{k+1}}^{(i)}$;
	\end{itemize}
	\item Compute $\hat p(\mathbf x_{t}|\mathbf y_{t_1:t_{k+1}})=\sum_{i=1}^N w_{t_{k+1}}^{(i)} \sum_{j=1}^M  \alpha(\mathbf{\tilde x}^{(i)(j)})\delta_{\mathbf{\tilde x}^{(i)(j)}}(\mathbf{ x}_t)$ for all $t \in ]t_{k},t_{k+1}]$.

\end{itemize}
\end{algorithm}

\section{One-dimensional simulation study}
\label{Simulation_study}

In this section, the smoothing method is experimented on a one-dimensional state space model. The results obtained with a standard particle-based smoothing are first presented in section \ref{Standard_smoothing_result}, and results of the proposed smoothing approach are shown in section \ref{Proposed_smoothing_result}

\subsection{State space model}
\begin{figure}[H]
\centering
\includegraphics[scale=0.5]{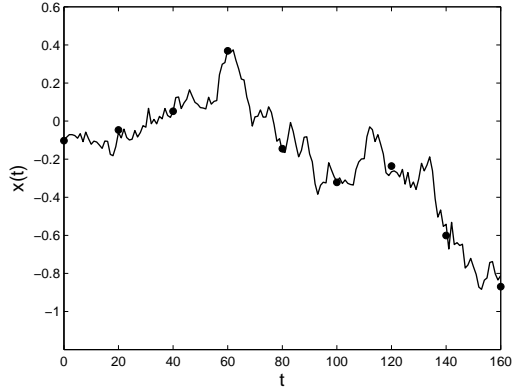}
\caption{Simulated sine diffusion trajectory $x(t)$ and partial observations $y(t_k)$ (dots) with $t_k-t_{k-1}=20\Delta t$. }
\label{fig:Sinus}
\end{figure}

The one-dimensional state space model of interest will be a sine diffusion, partially observed with noise (used as an illustration by \citet{Fearnhead08} for a particle filtering method) :
\begin{eqnarray}
{\rm d} x(t)&=&\sin(x(t)){\rm d}t + \sigma_x {\rm d}B(t),
\label{Sinus_state_space_model_1}\\
y_{t_k}&=&x_{t_k}+\gamma_{t_k},
\label{Sinus_state_space_model_2}
\end{eqnarray}
where $\sigma_x^2=0.5$ and $\gamma_{t_k} \sim \mathcal N(0,\sigma_y)$ with $\sigma_y^2=0.01$. One trajectory of the process is first simulated from (\ref{Sinus_state_space_model_1}) with an Euler-type discretization scheme of time step $\Delta t=0.005$. This trajectory will constitute the hidden process, observed through  $y_{t_k}$ generated according to (\ref{Sinus_state_space_model_2}) at every time step $t_k$, with $t_k-t_{k-1}=20\Delta t$. The trajectory is plotted on Figure \ref{fig:Sinus}, together with the corresponding discrete observations at times $t_k$.

~\\

\subsection{Standard fixed-lag smoothing}
\label{Standard_smoothing_result}

We present  the smoothing results obtained with the direct fixed-lag smoother presented in section \ref{Particles_smoothing}. Two situations are shown, with reduced and high number of particles. The case with a high number of particles is shown as the reference for comparison, note however that this ideal situation is not reachable in a high-dimensional context, since the number of particles has to be reduced for computational cost reasons.

Since the proposed method relies on a preliminary particle filtering step, filtering results are fist presented for the two situations: The first one is a particle filter with a small number of particles ($N=20$). The second case is a filter computed using $N=10000$ particles. The importance distribution that defines  this sequential importance sampling method is  chosen to be the transition law of the dynamic process (\ref{Sinus_state_space_model_1}). This is the standard choice for such a continuous-discrete filtering problem. 

\begin{figure}[H]
\begin{tabular}{cc}
   \includegraphics[scale=0.5]{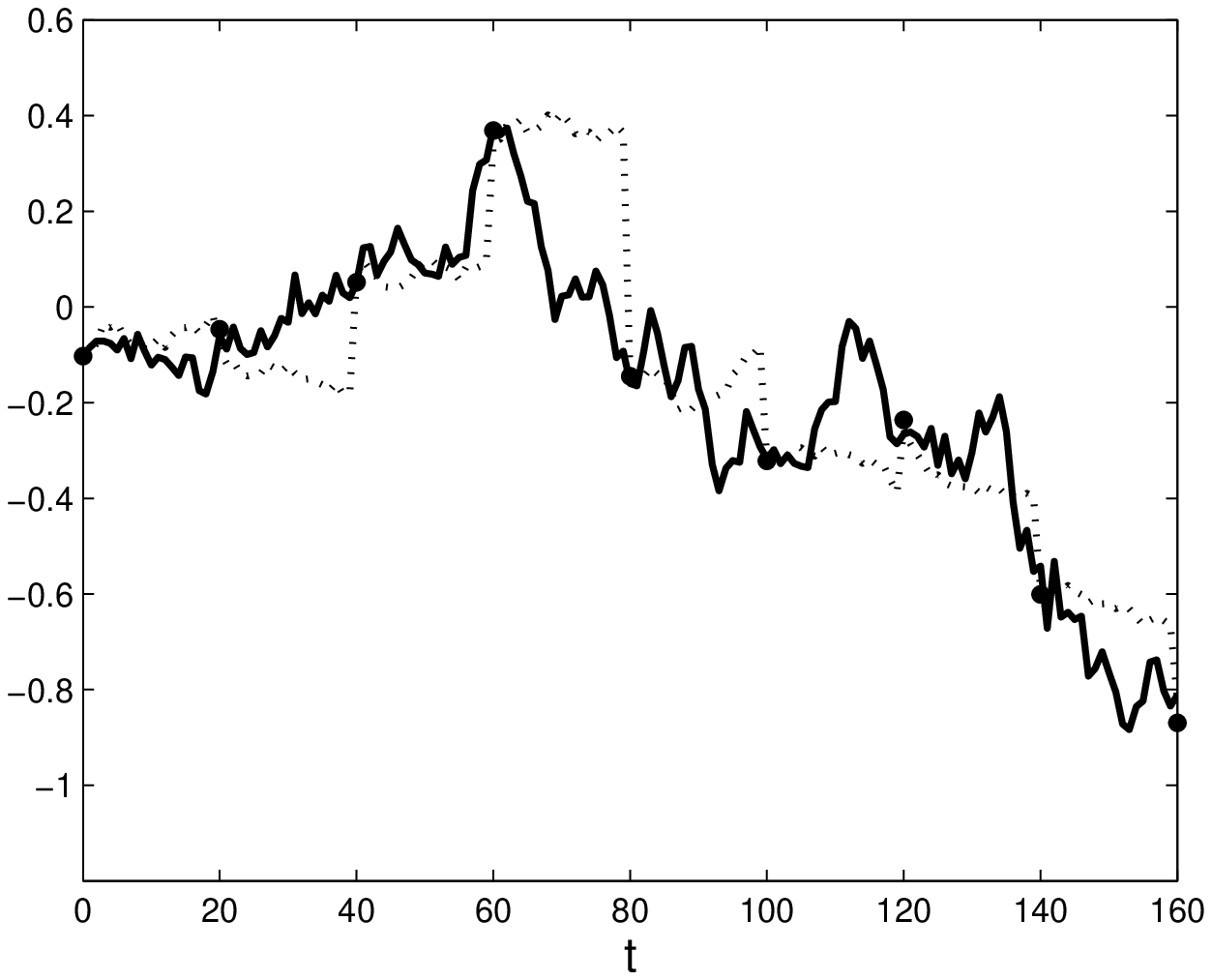} &
   \includegraphics[scale=0.5]{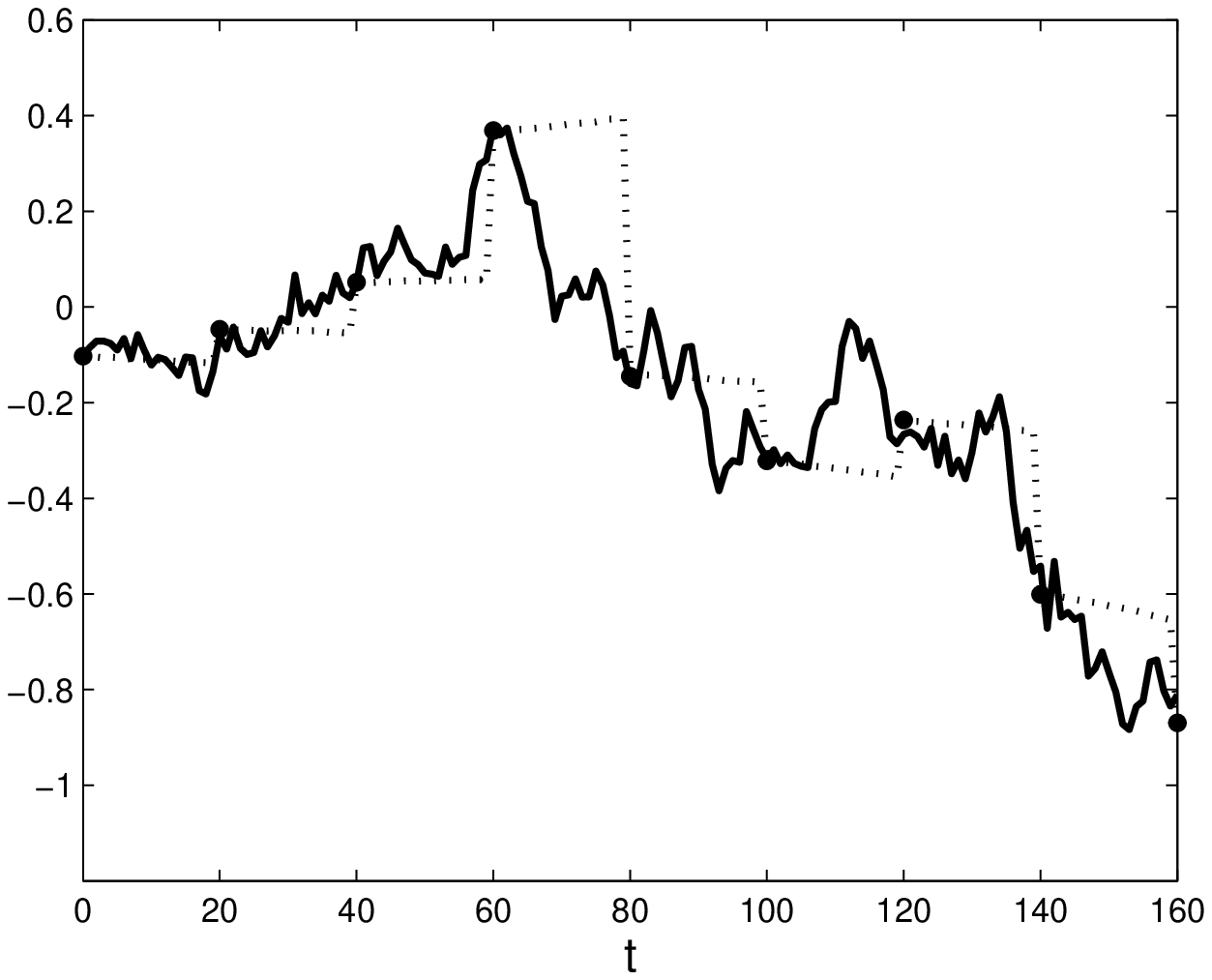} \\
    \footnotesize{(a)} & 
    \footnotesize{(b)} 
\end{tabular}
\caption{Particle filtering result. Thick line: hidden diffusion; Dots: partial observations; Dotted line: estimated filtering mean. (a) Result with $N=20$ particles. (b) Result with $N=10000$ particles.}
\label{fig:Filtrage_sinus}
\end{figure}

The  results for the two configurations are presented on Figure \ref{fig:Filtrage_sinus}, where the dotted lines represents the filtering mean estimates. The filtering distribution $p(x_{t_k}| y_{t_1:t_k})$ is estimated at each observation time $t_k$ using (\ref{Filtering_distribution_particles}), and predicted between observation times from (\ref{Predictive_filtering_distribution}). The mean is then estimated from weighted particles as $\sum_{i=1}^{N}w_{t_k}^{(i)}  x_{t}^{(i)}$, for all $t \in [t_k,t_{k+1}[$.
Figure \ref{fig:Filtrage_sinus} shows that both results (a) and (b) diverge from the reference solution between observation times. As a matter of fact, when no observation is available, the state distribution is predicted from the dynamics only, so that particles trajectories are not guided towards the next observation. At observation times $t_k$, high weights are given to particles that are close to the observation, so that the estimated  mean suddenly gets  closer to the solution. The fixed-lag smoothing approach implemented in the next section will aim at reducing the induced temporal discontinuities while providing dynamically consistent solutions.

~\\

\begin{figure}[H]
\begin{tabular}{cc}
   \includegraphics[scale=0.5]{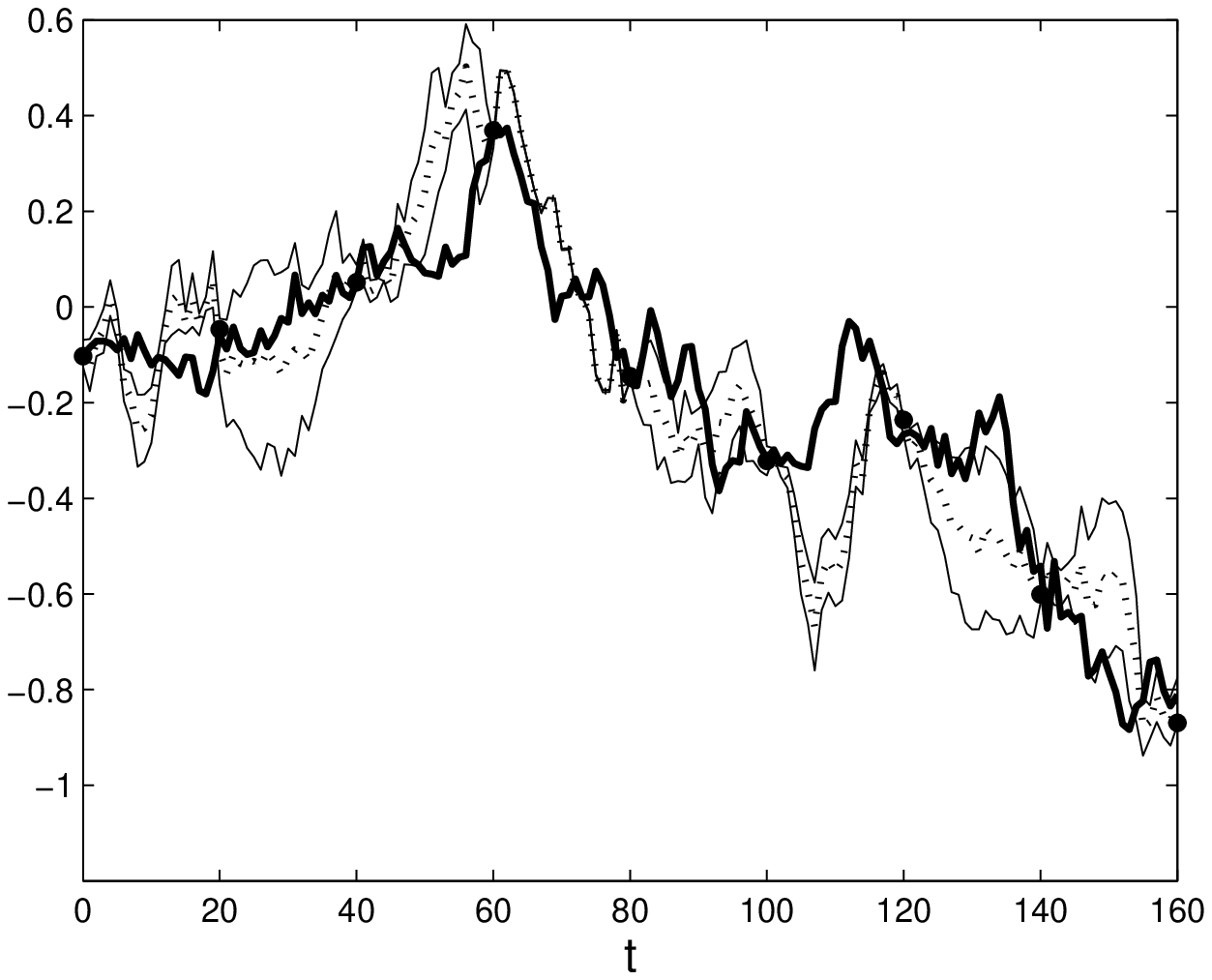} &
   \includegraphics[scale=0.5]{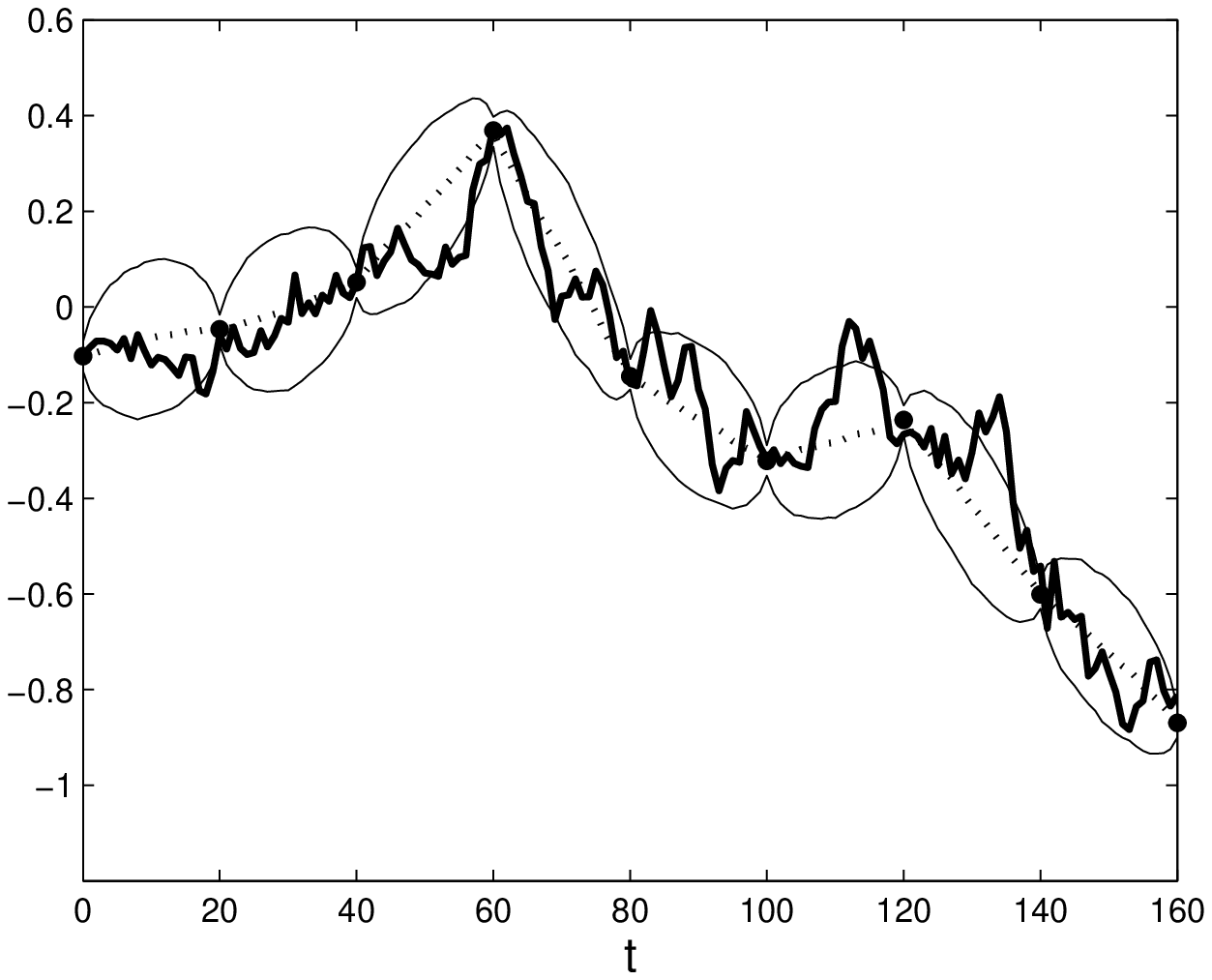} \\
    \footnotesize{(a)} & 
    \footnotesize{(b)} 
\end{tabular}
\caption{Standard particles fixed-lag smoothing result. Thick line: hidden diffusion; Dots: partial observations; Dotted line: estimated backward smoothing mean; Thin line: estimated standard deviation. (a) Result with $N=20$ particles. (b) Result with $N=10000$ particles.}
\label{fig:Smoothing_sinus}
\end{figure}

From this particle filtering result, we present now the results obtained from the direct particles smoothing procedure described in section \ref{Particles_smoothing} which relies on existing trajectories. The smoothing distribution $\hat{p}(x_{t}|y_{t_1:t_{k+1}})$ is computed backward for all $t \in ]t_{k},t_{k+1}]$ using expression (\ref{Local_smoothing_particles}), each time a new observation $y_{t_{k+1}}$ becomes available. Since this empirical distribution is computed from past trajectories that are reweighted with the new weights $w_{t_{k+1}}^{(i)}$ computed from the particle filter at time $t_{k+1}$, it can be poorly estimated if only a few weights are nonzero. This happens of course when the number $N$ of particles is too small so that only a few trajectories are close from the observation at time $t_{k+1}$.  This can be observed on Figure \ref{fig:Smoothing_sinus}(a), where the smoothing has been computed from the particle filtering result  with $N=20$ particles. The smoothing distribution $p(x_{t}|y_{t_1:t_{k+1}})$ is estimated using (\ref{Local_smoothing_particles}). The smoothing mean is computed as $\sum_{i=1}^{N}w_{t_{k+1}}^{(i)} x_{t}^{(i)}$ for all t $\in ]t_k,t_{k+1}]$, and the standard deviation is computed in the same way from the weighted particles. The mean is plotted with dotted line on Figure \ref{fig:Smoothing_sinus}, and the standard deviation envelope is plotted with thin line. We can note that at some  time intervals (for instance between observation times $t=100$ and $t=120$), the smoothing distribution is artificially peaked but far from the hidden trajectory. The smoothing result obtained from the particle filter with the reference case $N=10000$ particles is plotted on Figure \ref{fig:Smoothing_sinus}(b). In that configuration, since many trajectories have high weights at observation times,  the estimation of 
backward  smoothing distributions is improved and includes the hidden trajectory.

~\\

\subsection{Proposed smoothing}
\label{Proposed_smoothing_result}
In this section, we show how the proposed method can improve the estimation of backward smoothing distributions when it is not adequate to  rely on existing trajectories only. This is the case for instance if  the number of particles is too small, as demonstrated from the experiment presented on Figure  \ref{fig:Smoothing_sinus}.

\begin{figure}[H]
\begin{tabular}{cc}
   \includegraphics[scale=0.5]{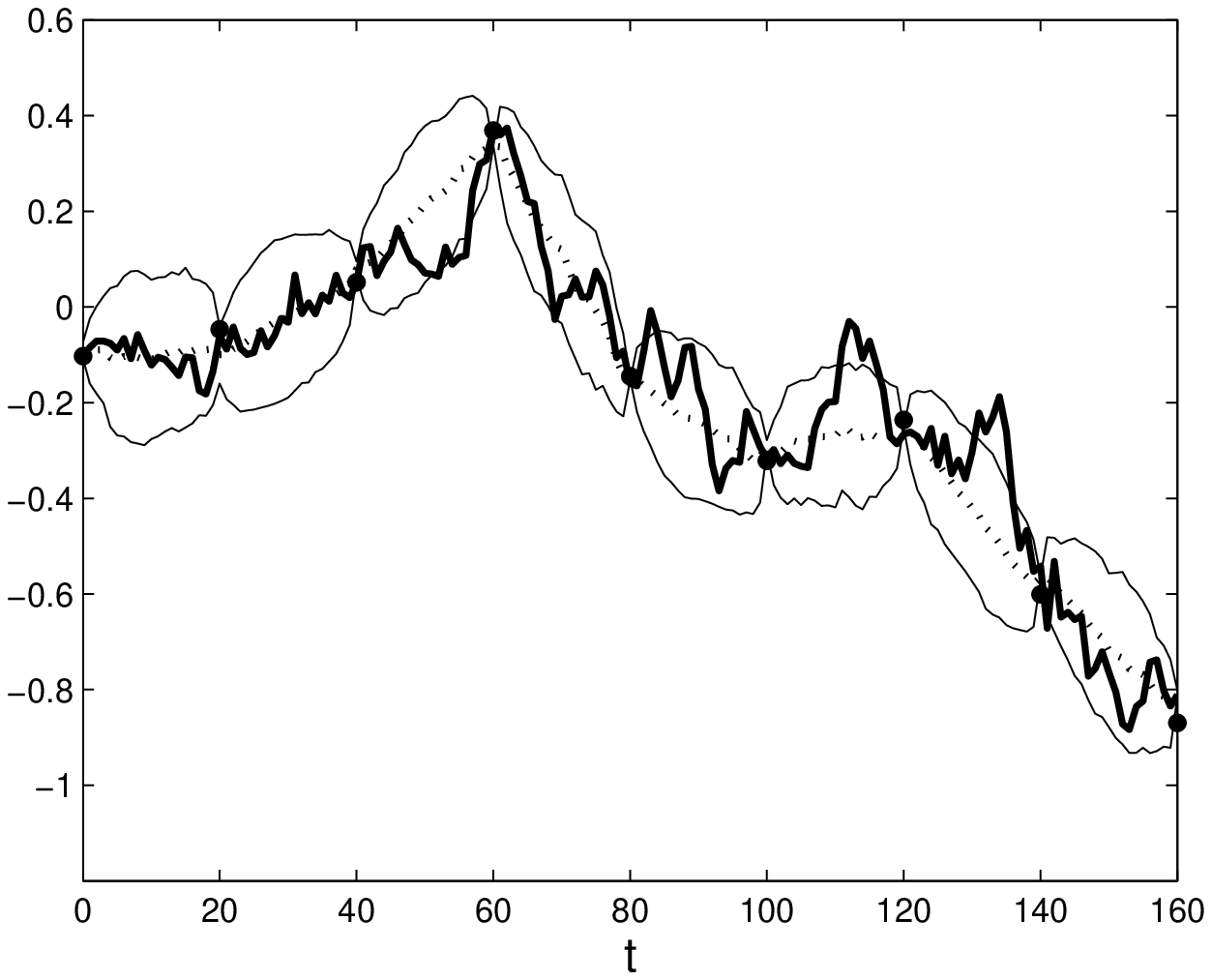} &
   \includegraphics[scale=0.5]{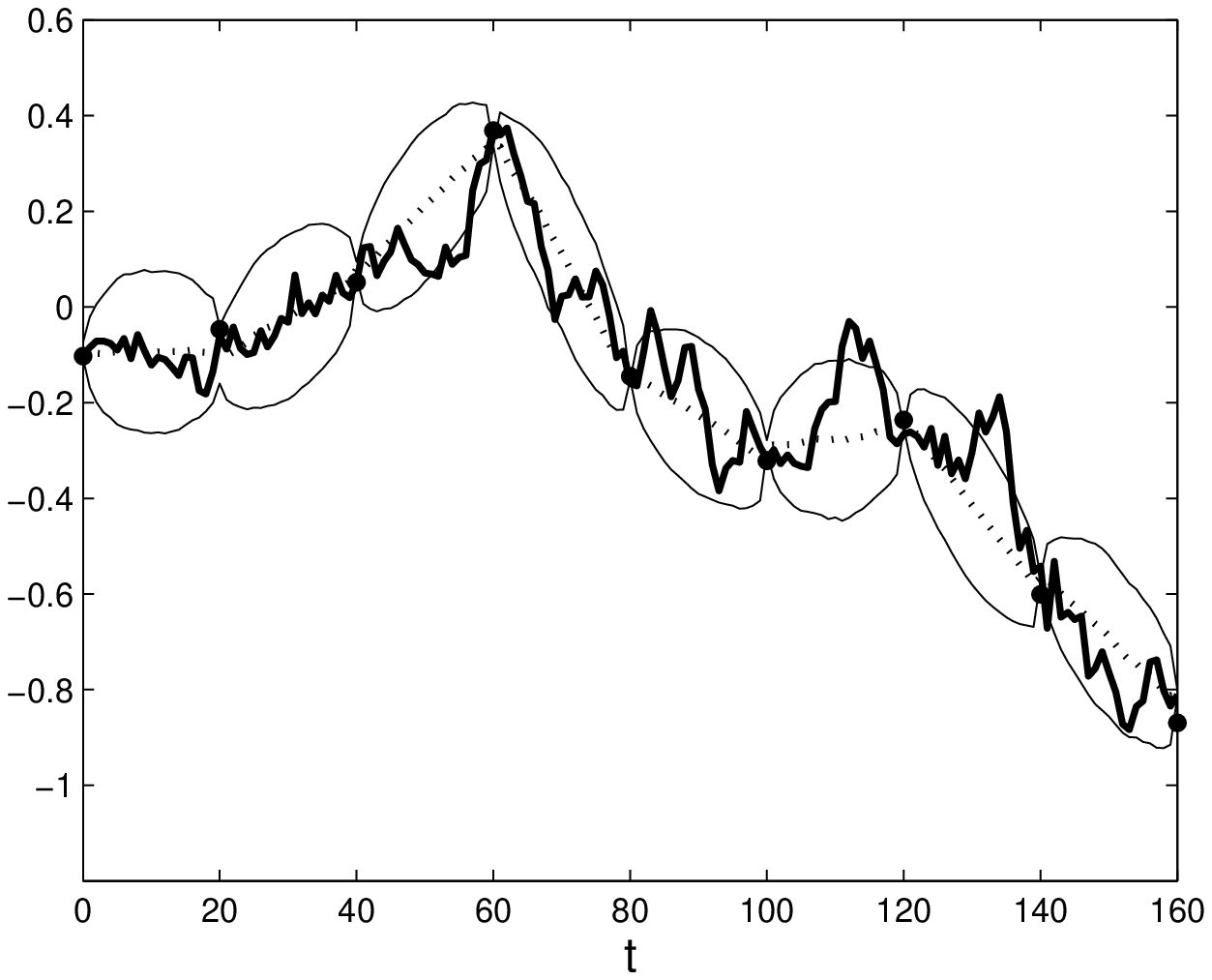} \\
    \footnotesize{(a)} & 
    \footnotesize{(b)} 
\end{tabular}
\caption{Proposed conditional smoothing result. Thick line: hidden diffusion; Dots: partial observations; Dotted line: estimated backward smoothing mean;  Thin line: estimated standard deviation. (a) Result with $N*M=20*50$ trajectories. (b) Result with $N*M=20*500$ trajectories.}
\label{fig:Smoothing_cond_sinus}
\end{figure}

Based on a particle filter result obtained with $N=20$ trajectories, Figure \ref{fig:Smoothing_cond_sinus}(a) shows the results obtained by our method  with $N*M=20*50$ trajectories, where we recall that $M$ is the number of conditional trajectories sampled between each pair  $\{x_{t_k}^{(i)},x_{t_{k+1}}^{(i)}\}$, $i=1,\ldots,N$. The smoothing distribution $\hat p(x_{t}|y_{t_1:t_{k+1}})$ is computed from (\ref{Conditional_smoothing_particles}), so the smoothing mean is computed as $\sum_{i=1}^N w_{t_k}^{(i)} \sum_{j=1}^M  \alpha(\tilde x^{(i)(j)}) \tilde x_t^{(i)(j)} $ for all $t \in ]t_k,t_{k+1}]$, and similarly for the standard deviation. The proposed method leads to improved smoothing distribution estimates in comparison to the direct particles smoothing approach presented on \ref{fig:Smoothing_sinus}(a). On Figure \ref{fig:Smoothing_cond_sinus}(b), the result obtained by the conditional smoothing technique is presented for $N*M=20*500$ trajectories. In that case, the result is very similar to the particles smoothing result presented on Figure \ref{fig:Smoothing_sinus}(b), obtained from a particle filter with $N=10000$. These results highlight the fact that since the proposed method creates new trajectories, it can improve the estimation of smoothing distributions when the initial number of filtering particles is too small.

~\\

\begin{figure}[H]
\begin{tabular}{cc}
   \includegraphics[scale=0.5]{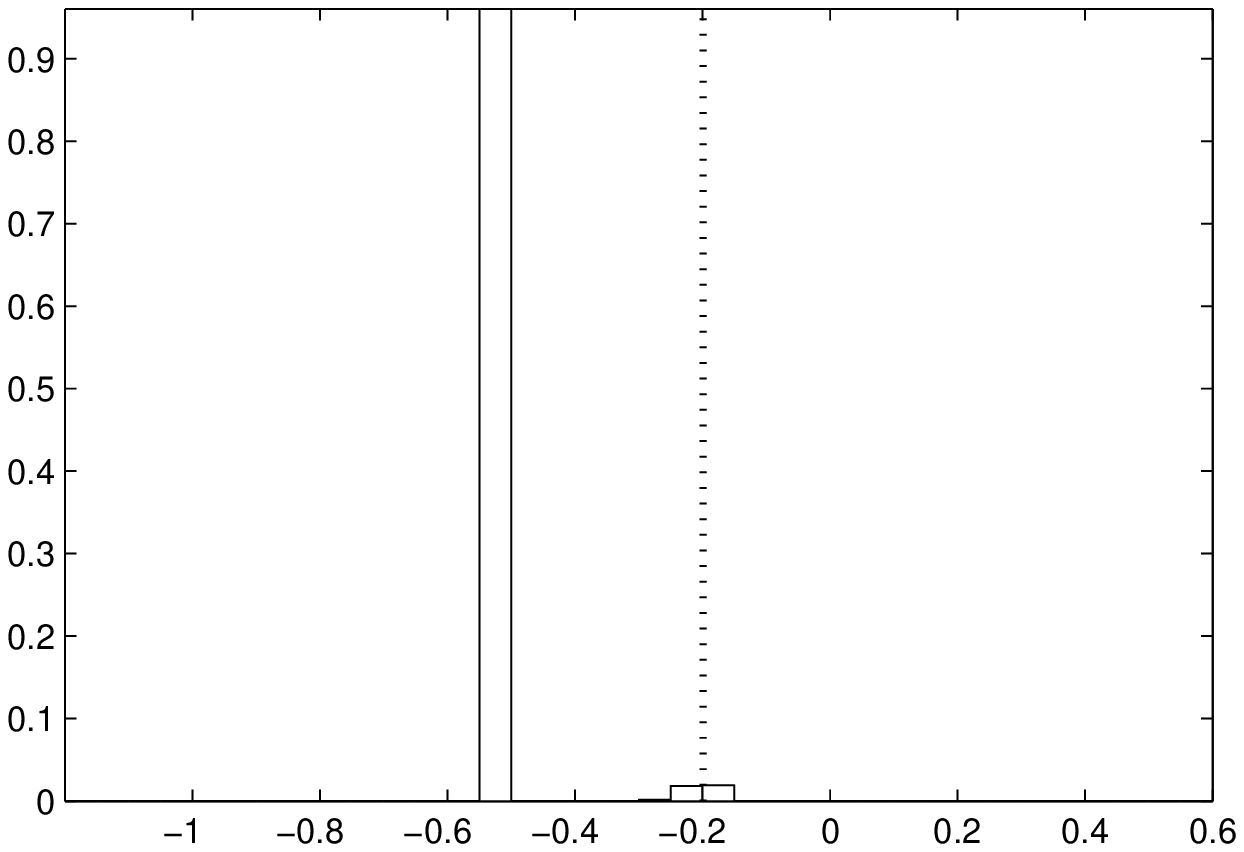} &
   \includegraphics[scale=0.5]{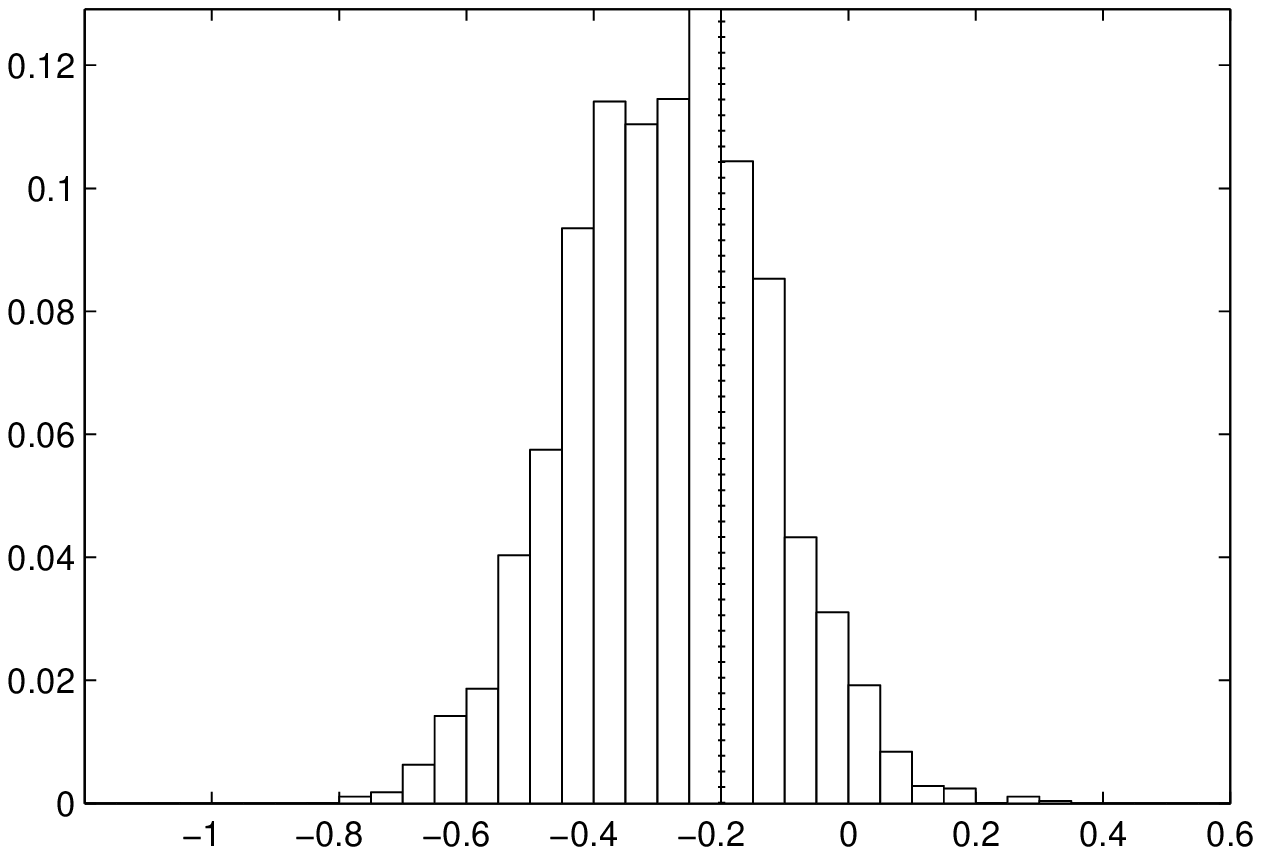} \\
     \footnotesize{(a)} & 
    \footnotesize{(b)} \\
   \includegraphics[scale=0.5]{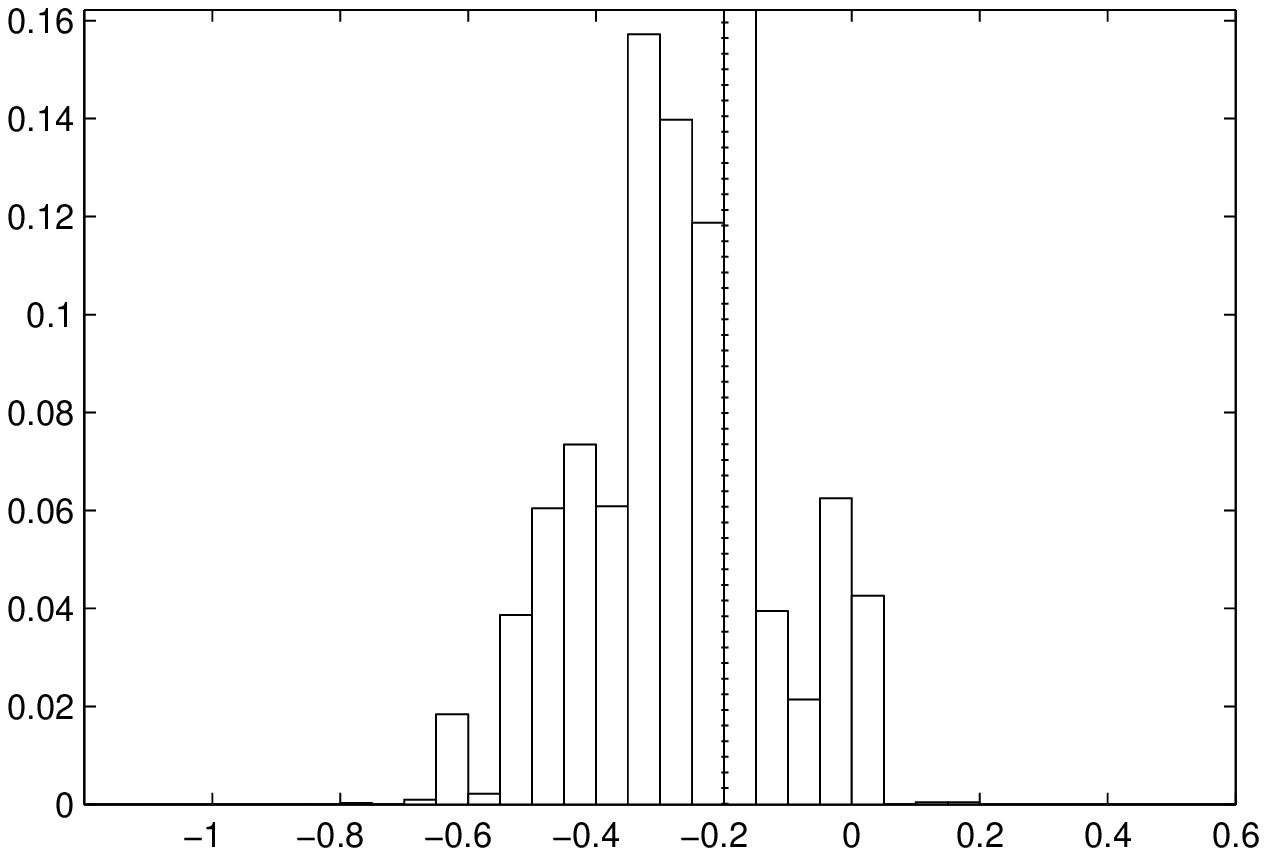} &
   \includegraphics[scale=0.5]{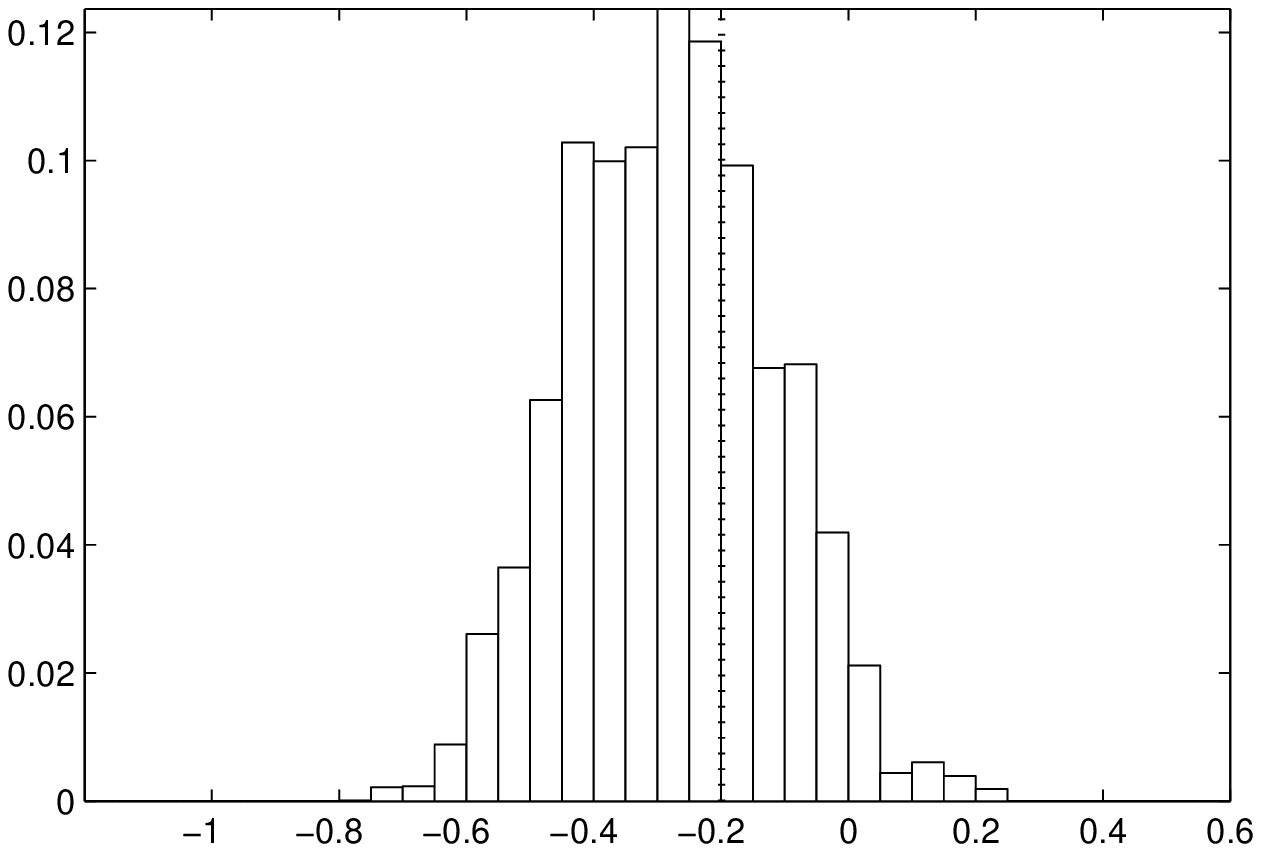} \\     
     \footnotesize{(c)} & 
    \footnotesize{(d)} \\
\end{tabular}
\caption{Estimated smoothing distributions $\hat p(x_{110}|y_{1:120})$. Dotted line: hidden known value $x_{110}$. (a) Particles smoothing results with $N=20$ particles. (b) Particles smoothing result with $N=10000$ particles. (c) Conditional smoothing result with $N*M=20*50$ trajectories. (d) Conditional smoothing result with $N*M=20*500$ trajectories.}
\label{fig:Hist_marginal_sinus}
\end{figure}

In addition, on Figure \ref{fig:Hist_marginal_sinus}, smoothing distributions are compared more precisely for a given time step ($t=110$) between  two observations at times $t_k=100$ and $t_{k+1}=120$. Histograms corresponding to the estimated smoothing distribution $\hat p(x_{110}|y_{1:120})$ are plotted for the particles  smoothing method with $N=20$ particles (Figure \ref{fig:Hist_marginal_sinus}(a)) and $N=10000$ particles (Figure \ref{fig:Hist_marginal_sinus}(b)), and the conditional smoothing method with $N*M=20*50$ trajectories (Figure \ref{fig:Hist_marginal_sinus}(c)) and $N*M=20*500$ trajectories (Figure \ref{fig:Hist_marginal_sinus}(d)). At this time step, the smoothing distribution based on $N=20$ particles is very peaked but not consistent with the hidden value (plotted as a dotted line). On the other hand, the support of the distribution obtained from the conditional method with $N*M=20*50$ trajectories is more consistent with the reference value. Moreover, as noted previously from Figure \ref{fig:Smoothing_cond_sinus}, the conditional smoothing solution with $N*M=20*500$ trajectories (Figure \ref{fig:Hist_marginal_sinus}(d)) is very similar to the solution obtained with a particle filter from $N=10000$ particles.

~\\

\section{Application to a high-dimensional  assimilation problem}
\label{Application}

This section aims at illustrating the applicability of our method to a high-dimensional and non linear scenario, without extensive study at this stage. The method is applied to a turbulence assimilation problem, where the state space model of interest is of type (\ref{State_space_model_1})-(\ref{State_space_model_2}). The goal is to recover temporal estimates of velocity/vorticity over a given spatial domain of size $n=64*64$, from a sequence of noisy observations and a continuous \textit{a priori} dynamical model based on a stochastic version of Navier-Stokes equation. Within an environmental framework, a direct application would be the estimation of wind fields or sea surface currents from satellite data. 

\subsection{State space model}
\label{State_space_fluid}

\begin{figure}[H]
\centering
\begin{tabular}{ccc}
   \includegraphics[width=3.5cm,height=3.5cm]{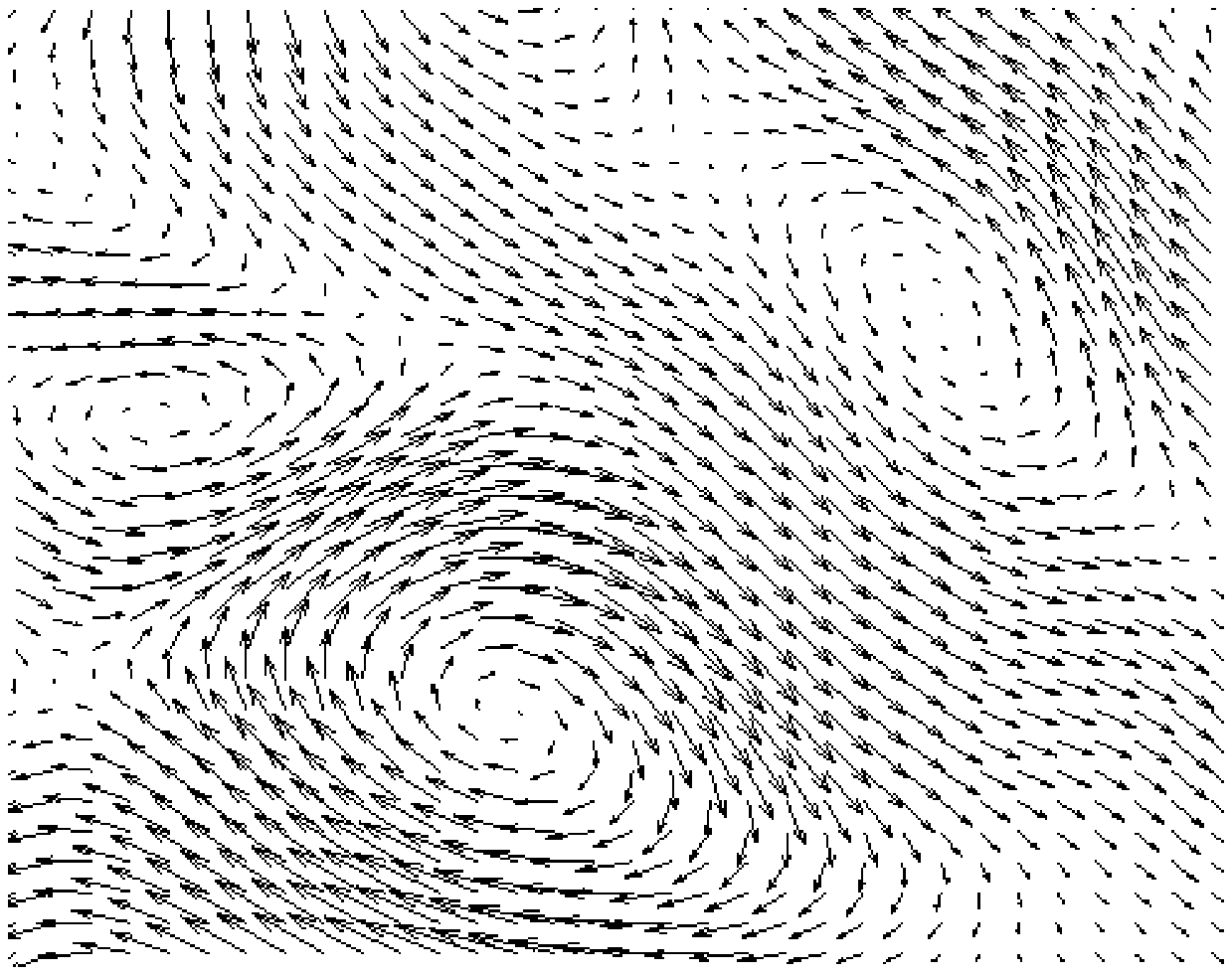} &
   \includegraphics[width=3.5cm,height=3.5cm]{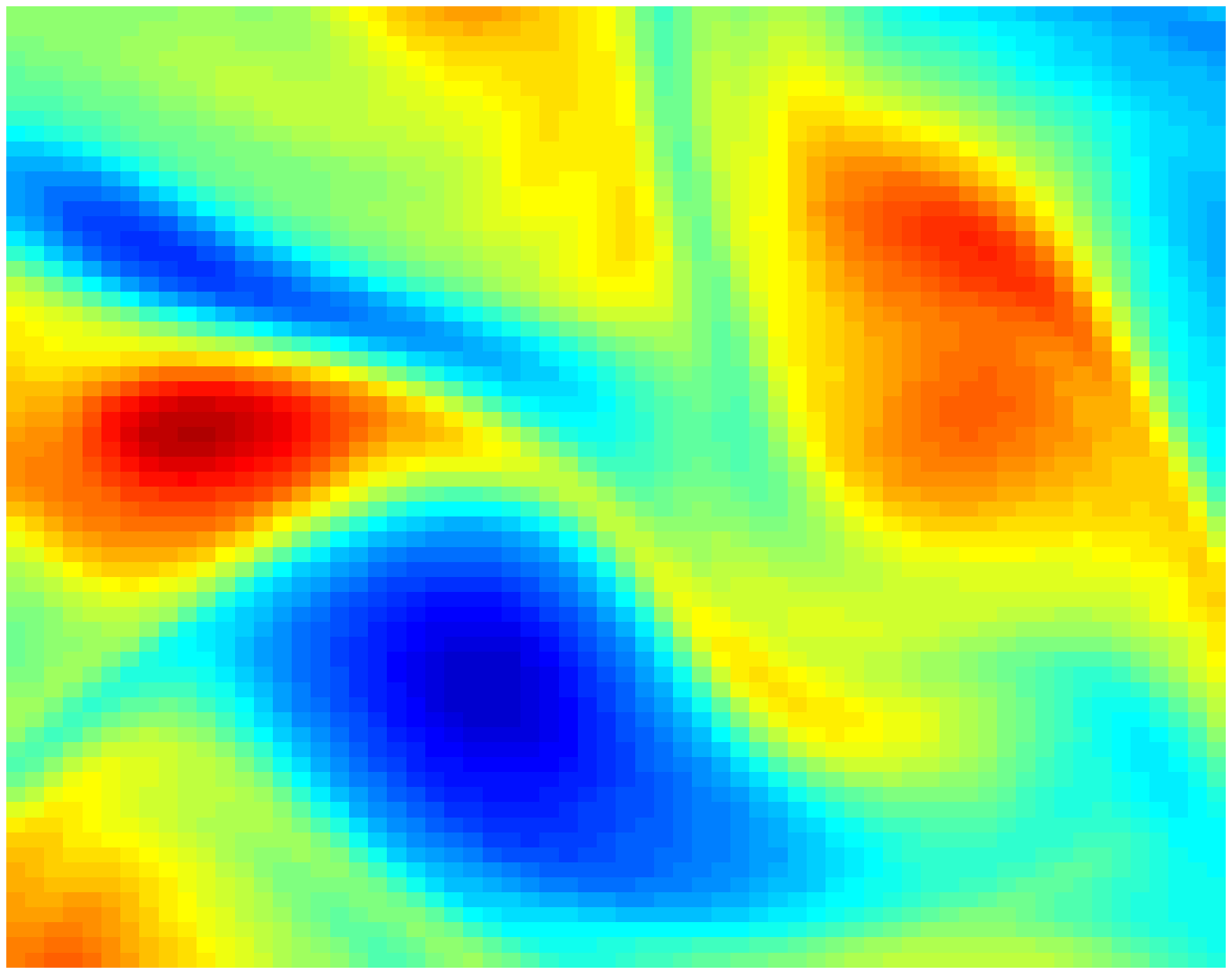}  &
   \includegraphics[width=0.5cm,height=3.5cm]{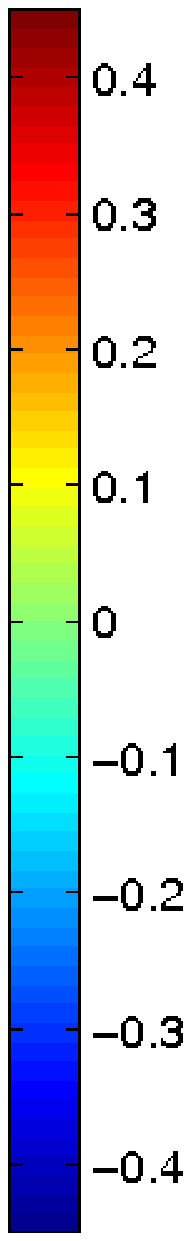}  
   \\
    \footnotesize{(a)} &     
    \footnotesize{(b)} & 
     \\
\end{tabular}
\caption{State example. (a) Velocity field $\mathbf w_t$; (b) Associated vorticity map $\boldsymbol \xi_t$.}
\label{fig:Groundtruth}
\end{figure}

Let $\xi(\mathbf x)$ denote the scalar vorticity at point $\mathbf x=(x,y)^T$, associated to the 2D velocity $w(\mathbf x)=(w_x(\mathbf x),w_y(\mathbf x))^T$ through $\xi(\mathbf x)=\frac{\partial w_y}{\partial x}-\frac{\partial w_x}{\partial y}$. Let $\boldsymbol \xi \in \mathbb R^n$ be the state vector describing the vorticity over a $n=64*64$ square domain, and $\mathbf w \in \mathbb R^{2n}$ the associated velocity field over the domain. We will focus on incompressible flows such that the divergence of the velocity field is null. A stochastic version of Navier-Stokes equation in its velocity-vorticity form can then be written as:
\begin{equation}
{\rm d}\boldsymbol \xi_t = -  \nabla  \boldsymbol \xi_t \cdot \mathbf{w}_t {\rm d}t +  \nu  \Delta \boldsymbol \xi_t {\rm d}t + \sigma {\rm d}\mathbf B_t,
\label{Navier_Stokes_stochastic}
\end{equation}
where $\nu$ denotes the fluid viscosity coefficient (assumed to be known). The uncertainty is modeled by a Brownian motion of size $n$, with covariance $\Sigma=\sigma\sigma^T$, where $\sigma \in \mathbb R^n$. A velocity field example, generated from the model  (\ref{Navier_Stokes_stochastic}), is shown on Figure \ref{fig:Groundtruth}(a), together with the corresponding vorticity map (b).

We assume the hidden vorticity vector $\boldsymbol \xi$ is observed through noisy measurements $\mathbf y_{t_k}$ at discrete times $t_k$, where $t_k-t_{k-1}=100 \Delta t$, and $\Delta t=0.1$ is the time step used to discretize (\ref{Navier_Stokes_stochastic}). In our experimental setup, measurements correspond to PIV (Particle Image Velocimetry) image sequences used in fluid mechanics applications. Note however that other kind of data can be used similarly within this state space model, like meteorological or oceanographic data for instance. The state and observation are  related in our case through $\mathbf y_{t_k}=g(\boldsymbol \xi_{t_k})+\gamma_{t_k}$, where $g$ is a non linear function linking the vorticity to the image data, and $\gamma_{t_k}$ is a Gaussian noise, uncorrelated in time.

~\\

\subsection{Implementation details}
We recall that the smoothing  relies first on a particle filter step. Due to the high dimensionality of the state vector, the use of a standard particle filter is not adapted to solve the filtering problem, as discussed by \citet{Snyder08} or \citet{VanLeeuwen09}. We make then use of the method presented by \citet{Papadakis10} which combines the benefits of the ensemble Kalman filter,  known to perform well in practice for high dimensional systems \citep{Stroud10}, and the particle filter (which solves theoretically the true filtering problem, without approximating the filtering distributions with Gaussian distributions). Since the method of \citet{Papadakis10} is  intrinsically a particle filter, it leads then at each observation time $t_k$ to a set of  particles and weights $\{\boldsymbol \xi^{(i)}_{t_1:t_k},w_{t_k}\}_{i=1:N}$, as required by the algorithm proposed in section \ref{Conditional_smoothing}.

The particle filter step requires simulations from the dynamical model (\ref{Navier_Stokes_stochastic}), and the conditional simulation step requires to sample trajectories from its constrained version, which consists in a similar problem with modified drift (see  process (\ref{Diffusion_process_constraint})). The model is discretized in time with time step $\Delta t=0.1$; more information about the discretization scheme may be obtained in \citet{Papadakis10}. The random perturbations are assumed to be realizations of Gaussian random fields that are correlated in space with exponential covariance structure 
$\Sigma(\mathbf x_i,\mathbf x_j)=\eta\exp(-\frac{||\mathbf x_i-\mathbf x_j||^2}{\lambda})$, where $\eta=0.01$ and $\lambda=13$. In practice, the simulation of these perturbations is performed in Fourier space, with the method described in \citet{Evensen03}.

Finally, the estimation of the smoothing distributions require the computation of conditional trajectories weights, corresponding to Girsanov weights given by (\ref{Girsanov_weights}). After a Riemann sum approximation of the integral, the computation of weights requires the inversion of the matrix $\Sigma$ of size $(n,n)$, where $n=64*64$ is the number of grid points. We choose to compute $\Sigma^{-1}$ empirically using a singular value decomposition computed from the $M$ realizations of the perturbation fields used for the constrained trajectories simulations. Let $\mathbf Z$ be the matrix of size $(n,M)$ containing the $M$ centered fields of size $n=64*64$, the SVD leads to $\mathbf Z=\mathbf U \mathbf D \mathbf V^T$ , so that $\mathbf Z\mathbf Z^T=\mathbf U \mathbf D \mathbf D^T\mathbf U^T$.  The inverse of the covariance matrix $\Sigma^{-1}$ is finally computed as:
\begin{equation}
 M(\mathbf Z\mathbf Z^T)^{-1}=M\mathbf U (\mathbf D \mathbf D^T)^{-1}\mathbf U^T,
\end{equation}
which only requires the inversion of a diagonal.

\subsection{Results}
In this section, we illustrate the capability of the proposed method to reduce the temporal discontinuities inherent to the particle filter in our continuous-discrete state-space  setting.

\begin{figure}[H]
\centering
\includegraphics[width=12cm,height=5cm]{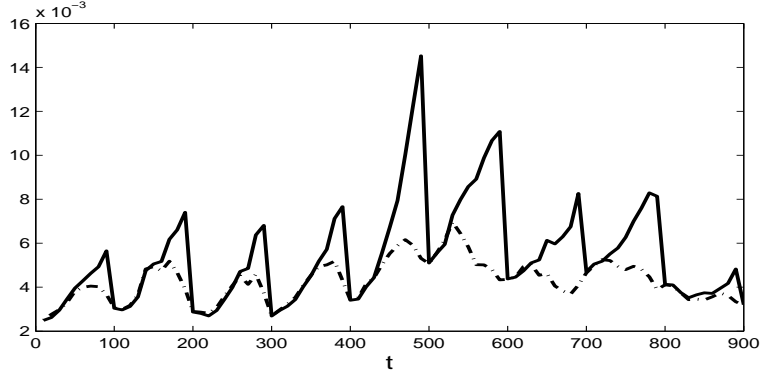}
\caption{Full line: mean square error between ground truth vorticity and estimated filtering mean; Dotted line: mean square error between ground truth vorticity and estimated backward smoothing mean.}
\label{fig:Vorti_error}
\end{figure}

The particle filter step has been computed from $N=500$ particles. Since the ground truth vorticity sequence is known in our experimental setup, the mean square error can be computed between the hidden vorticity and the estimated filtering mean, given by $\sum_{i=1}^{N}w_{t_k}^{(i)} \boldsymbol \xi_{t}^{(i)}$ for all $t \in [t_k,t_{k+1}[$. This error, averaged over the  domain of size $n=64*64$, is plotted on Figure \ref{fig:Vorti_error} with full line. As observed in section \ref{Simulation_study} for the one-dimensional example, the correction of the filtering solution at observation times leads to  sudden error decreases.
The proposed smoothing method has been applied with $M=200$. In practice, many filtering trajectories have close to zero weights at observation times (note however that the filter is not degenerate and is able to recover the hidden vorticity, as shows the filtering result presented on Figure \ref{fig:Vorti_error}). This implies that the method relies in practice on a reduced number $\tilde NM$ of sampled conditional trajectories (with $\tilde N << N$), which makes the problem computationally tractable. The smoothing distribution $\hat p(\boldsymbol \xi_{t}|\mathbf y_{t_1:t_{k+1}})$ is computed for all $t \in ]t_k,t_{k+1}]$ from (\ref{Conditional_smoothing_particles}), and its mean is computed as $\sum_{i=1}^N w_{t_{k+1}}^{(i)} \sum_{j=1}^M  \alpha(\tilde{\boldsymbol \xi}^{(i)(j)}) \tilde{\boldsymbol \xi}_{t}^{(i)(j)}$. The mean square error is computed between the true vorticity and the estimated smoothing mean, and plotted on Figure \ref{fig:Vorti_error} with dotted line. As expected, the smoothing method reduces the error at hidden times between observations.

In addition, we present below a qualitative evaluation of the smoothing result for the same experiment, over a specific time interval.  

\begin{figure}[H]
\centering
\begin{tabular}{cc}
\begin{tabular}{ccc}
   \includegraphics[width=3.5cm,height=3.5cm]{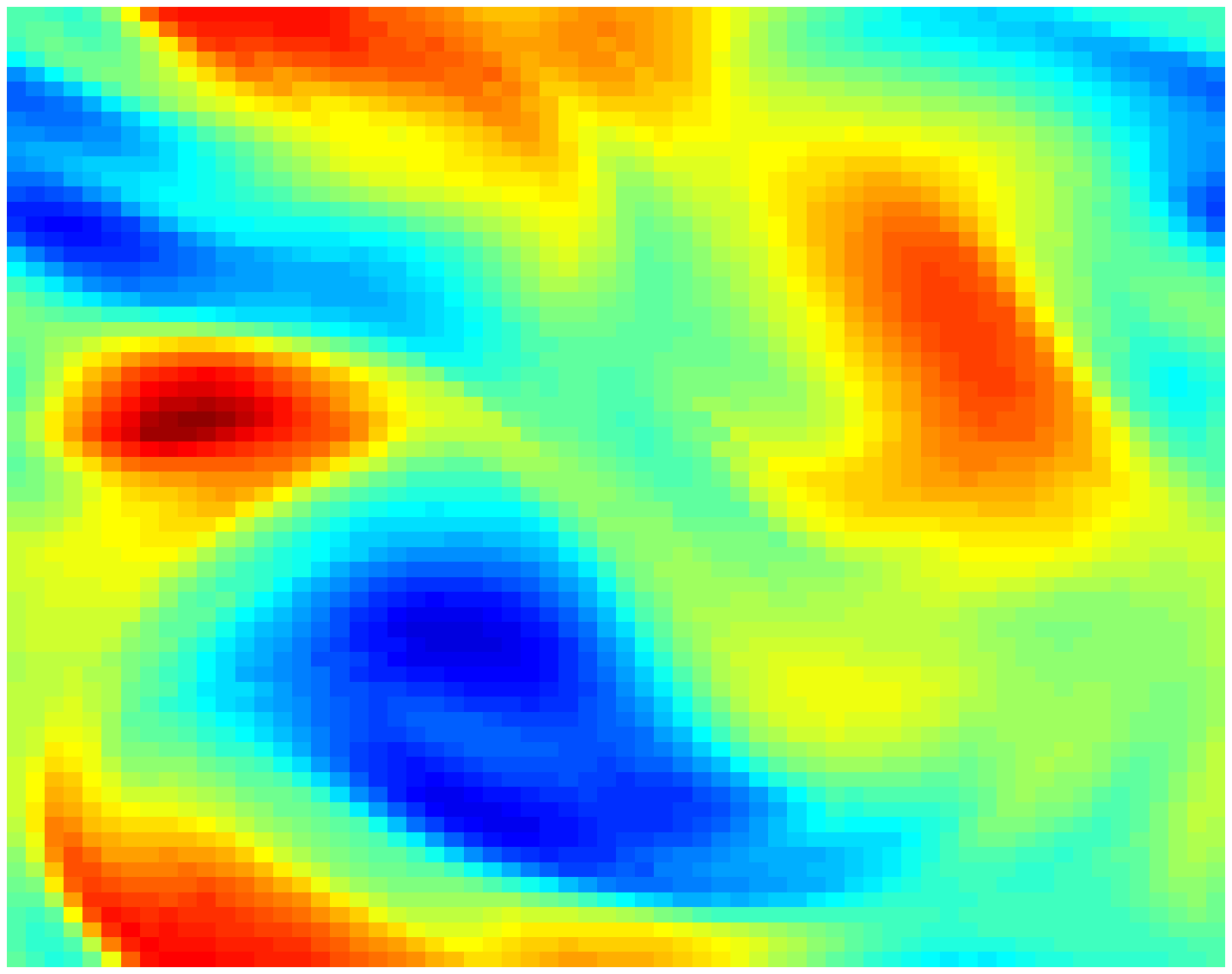} &
   \includegraphics[width=3.5cm,height=3.5cm]{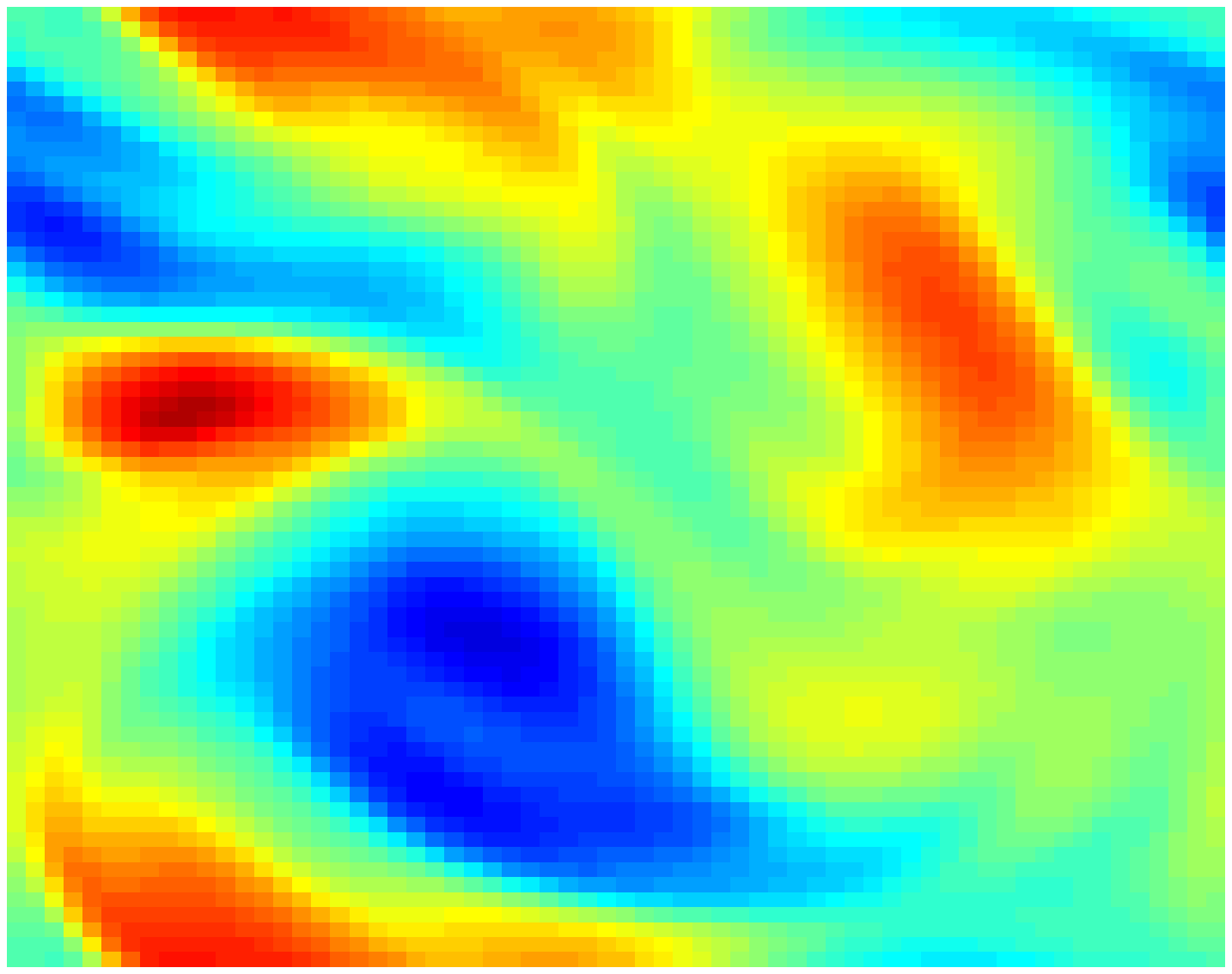} &
   \includegraphics[width=3.5cm,height=3.5cm]{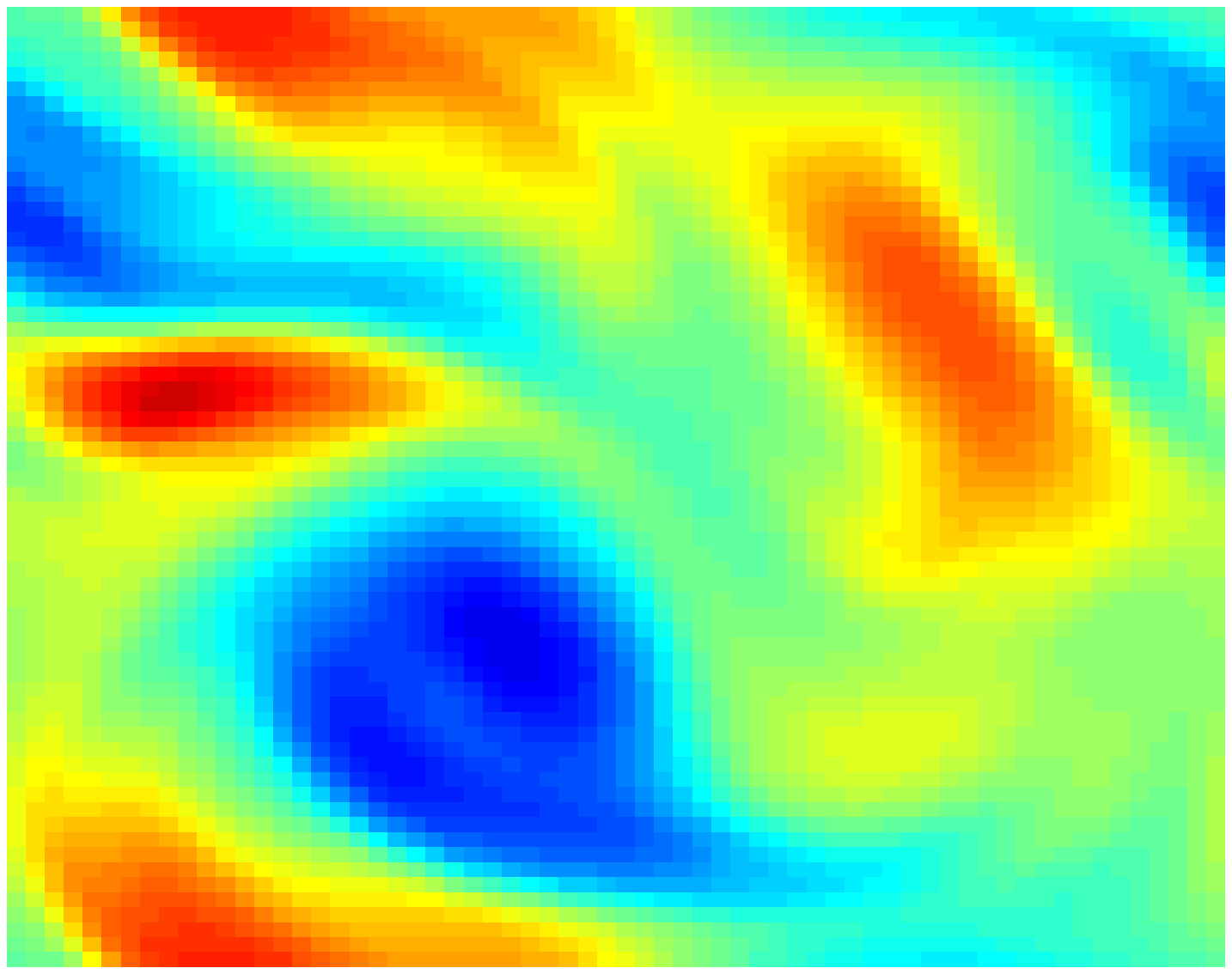} \\
     \footnotesize{$t=400$} & 
     \footnotesize{$t=420$} & 
     \footnotesize{$t=450$} \\
   \includegraphics[width=3.5cm,height=3.5cm]{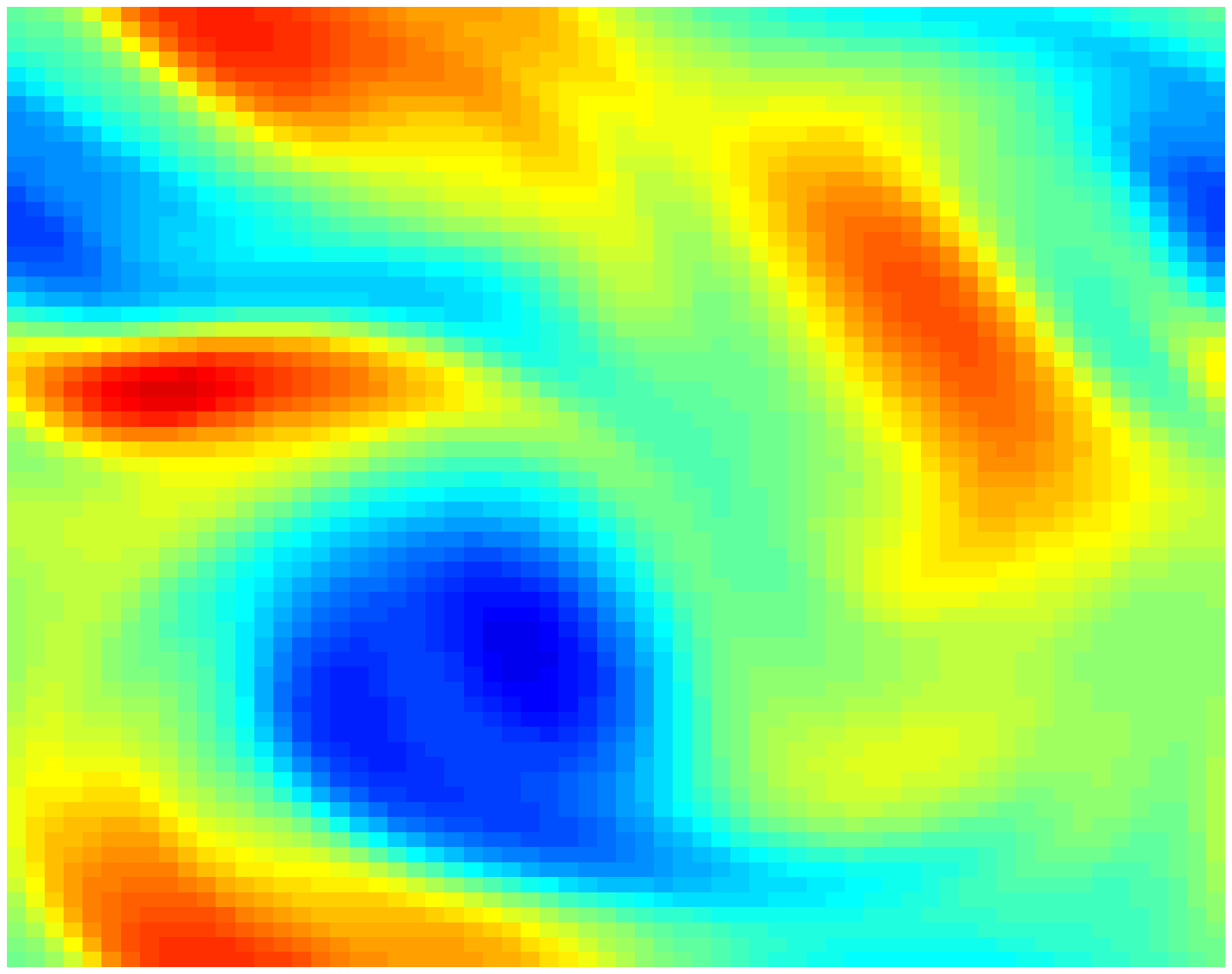} &
   \includegraphics[width=3.5cm,height=3.5cm]{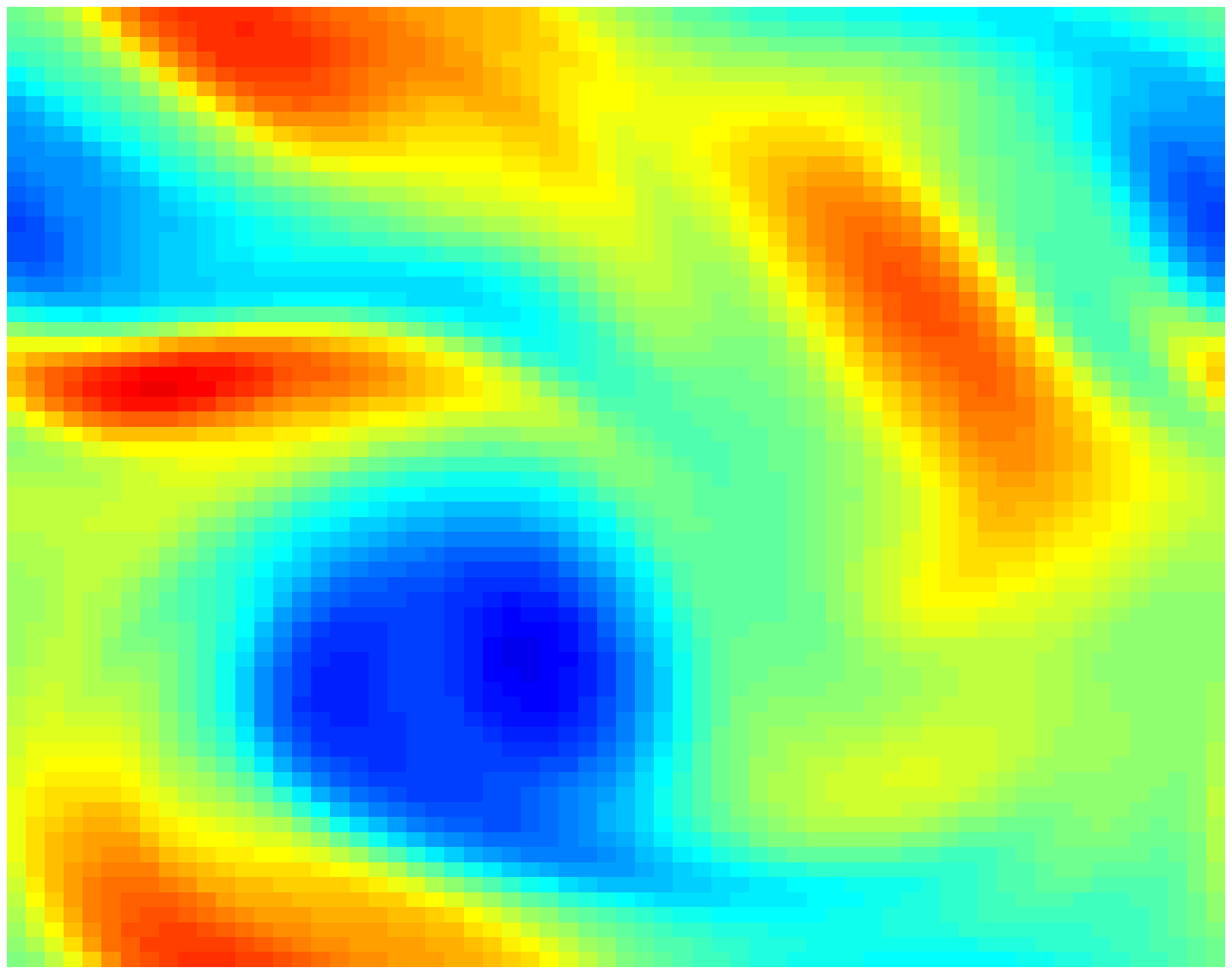} &
   \includegraphics[width=3.5cm,height=3.5cm]{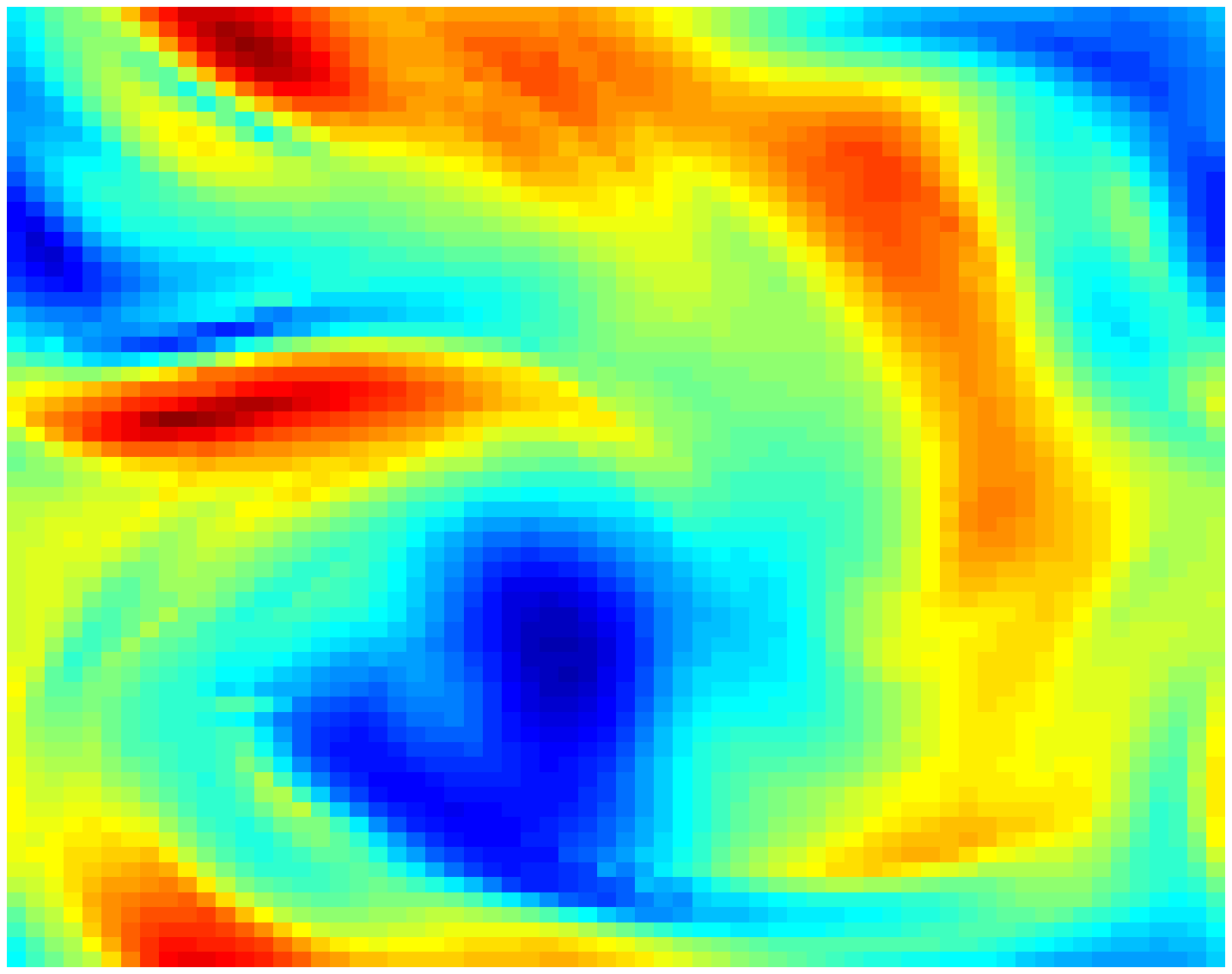} \\
 \footnotesize{$t=470$} & 
     \footnotesize{$t=490$} & 
    \footnotesize{$t=500$} \\
    \end{tabular}
    &
     \begin{tabular}[]{c} 
\includegraphics[width=0.5cm,height=3.5cm]{Scale_vorti.eps} 
\end{tabular}
    \end{tabular}\caption{Filtering result with the method of \citet{Papadakis10}. Estimated  mean vorticity maps for different times $t$ between observation times $t=400$ and $t=500$.}
\label{fig:Vorti_filtering}
\end{figure}

\begin{figure}[H]
\centering
\begin{tabular}{cc}
\begin{tabular}{ccc}
   \includegraphics[width=3.5cm,height=3.5cm]{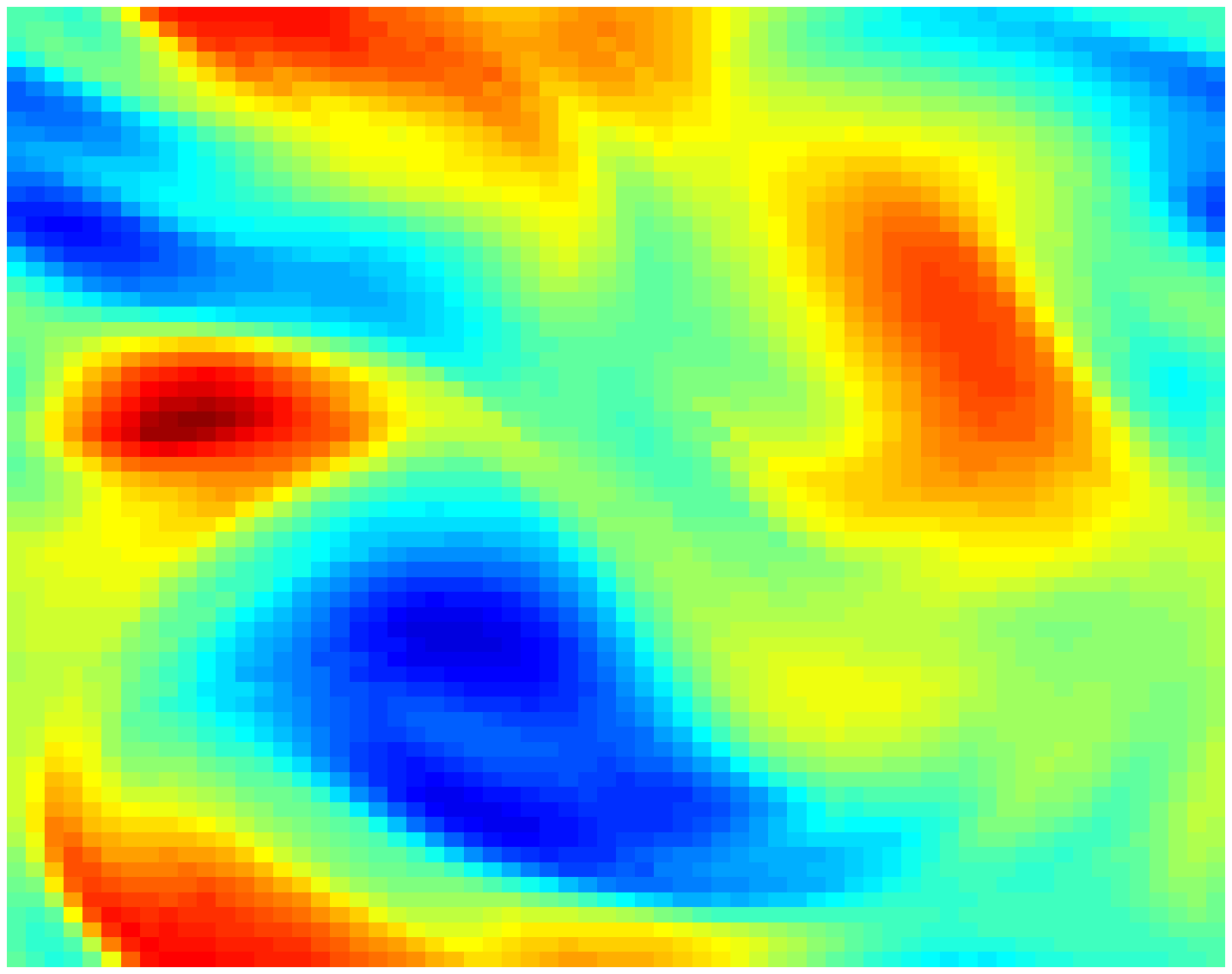} &
   \includegraphics[width=3.5cm,height=3.5cm]{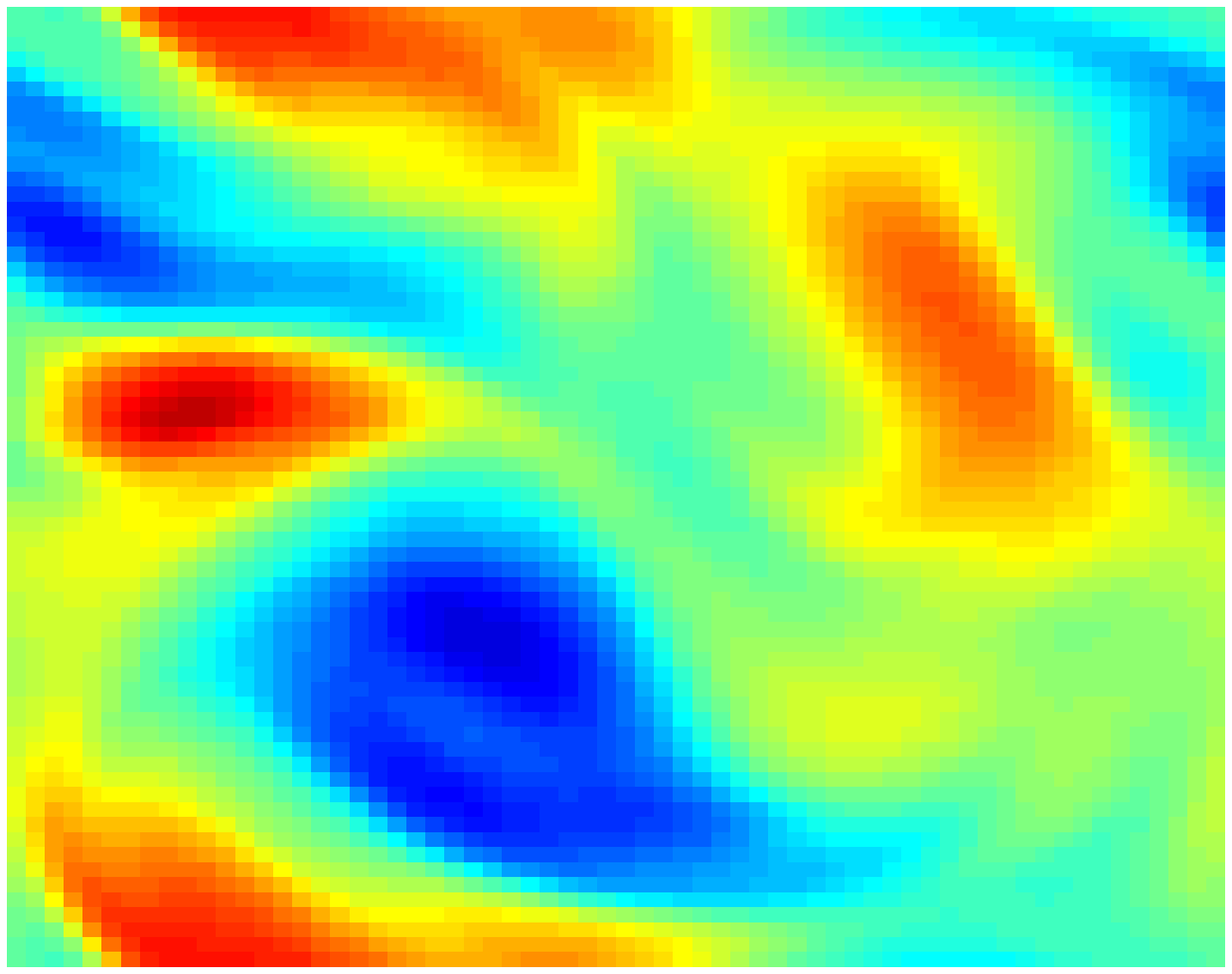} &
   \includegraphics[width=3.5cm,height=3.5cm]{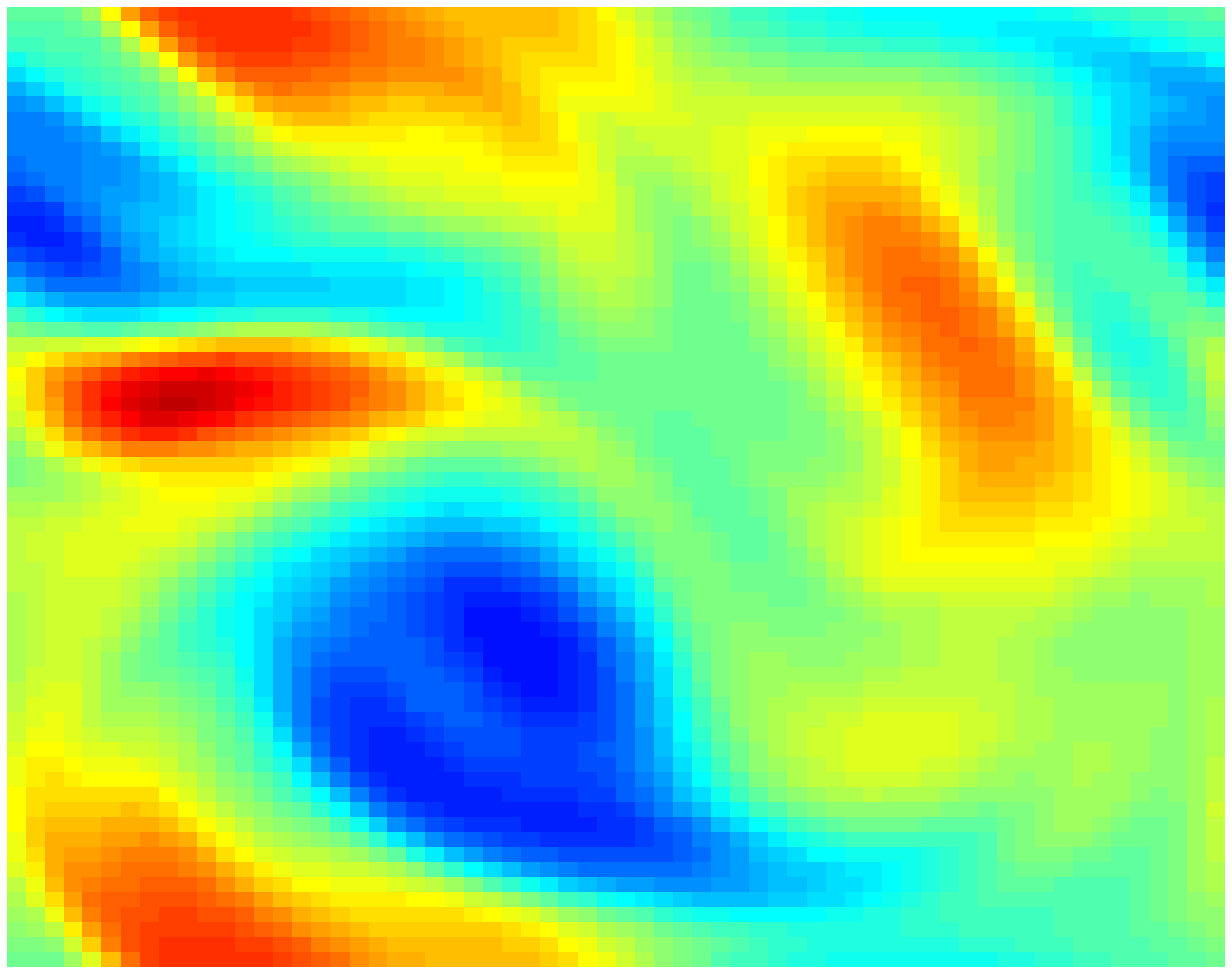} \\
     \footnotesize{$t=400$} & 
     \footnotesize{$t=420$} & 
     \footnotesize{$t=450$} \\
   \includegraphics[width=3.5cm,height=3.5cm]{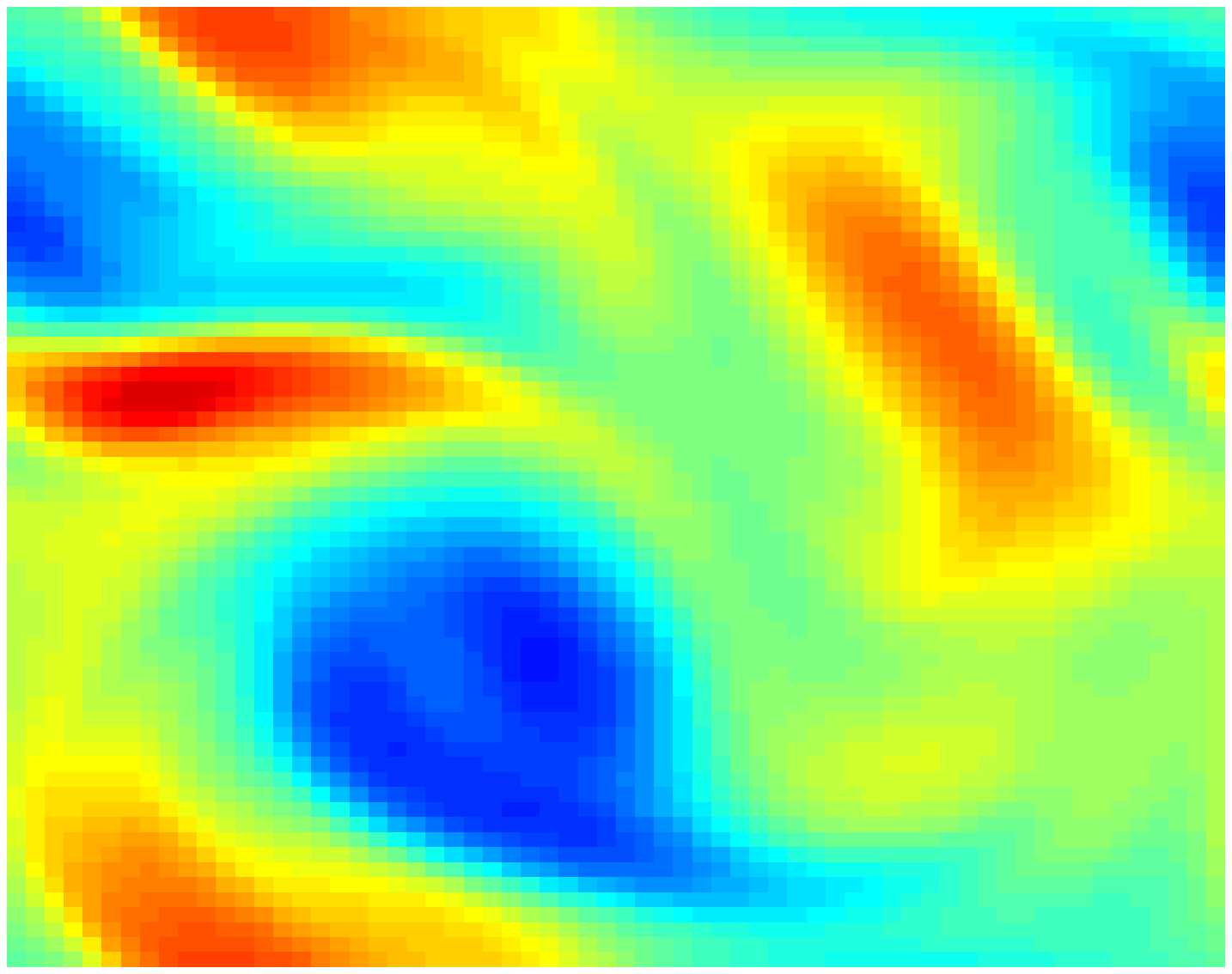} &
   \includegraphics[width=3.5cm,height=3.5cm]{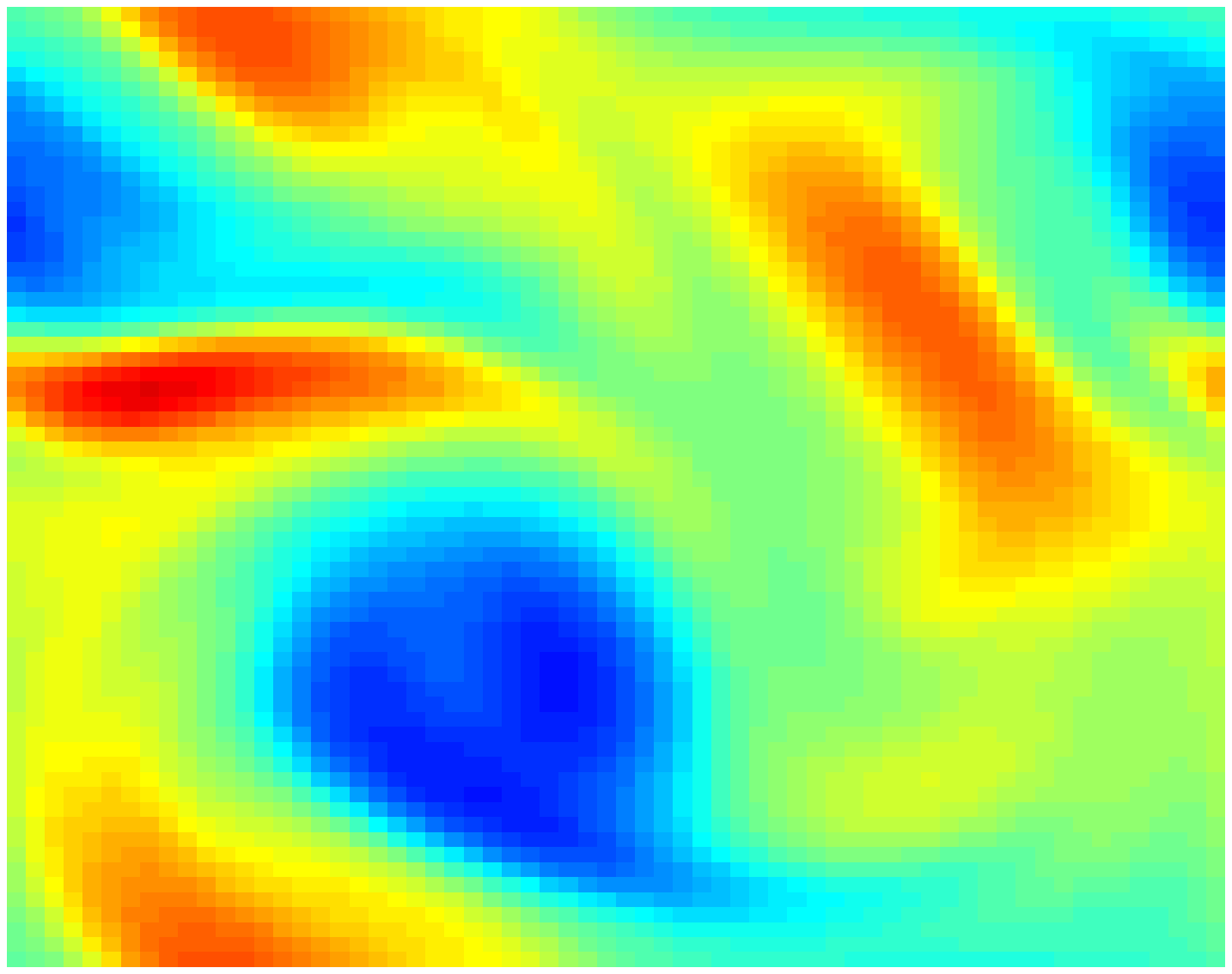} &
   \includegraphics[width=3.5cm,height=3.5cm]{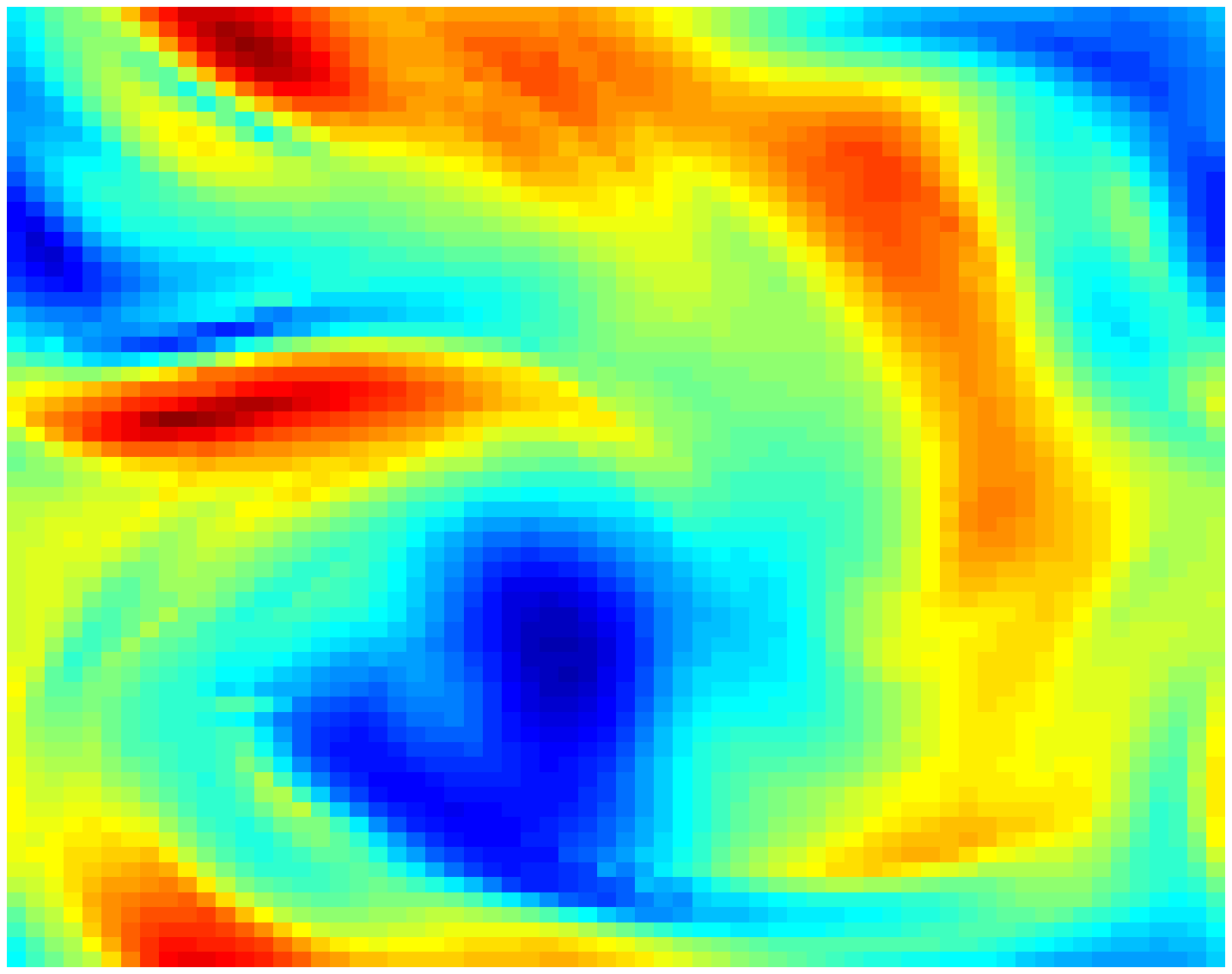} \\
 \footnotesize{$t=470$} & 
     \footnotesize{$t=490$} & 
    \footnotesize{$t=500$} \\
    \end{tabular}
    &
     \begin{tabular}[]{c} 
\includegraphics[width=0.5cm,height=3.5cm]{Scale_vorti.eps} 
\end{tabular}
    \end{tabular}
\caption{Standard particles smoothing result (see Section \ref{Particles_smoothing}). Estimated  mean vorticity maps for different times $t$ between observation times $t=400$ and $t=500$.}
\label{fig:Vorti_standard_smoothing}
\end{figure}

\noindent The particle filter result is  first presented on Figure \ref{fig:Vorti_filtering} for the time interval $[400,500]$ between two observations, where estimated mean vorticity maps are computed as $\sum_{i=1}^{N}w_{400}^{(i)} \boldsymbol \xi_{t}^{(i)}$ for all $t \in [400,500[$, and as $\sum_{i=1}^{N}w_{500}^{(i)} \boldsymbol \xi_{t}^{(i)}$ for $t=500$. The temporal discontinuity between estimations can be observed when reaching  observation time $t=500$: the vorticity map is suddenly modified in order to fit to the observations, introducing  inconsistencies in the vorticity temporal trajectories. Note that the application of the standard particles smoothing (described in section \ref{Particles_smoothing}) will fail here, and not only because the number of particles is too small. As a matter of fact, we recall that the filtering trajectories have been computed from the method presented in \citet{Papadakis10}, which uses the ensemble Kalman filter step as importance distribution in the particle filter algorithm. The ensemble Kalman filter consists of a prediction step from the dynamical model (\ref{Navier_Stokes_stochastic}), and a correction step which shifts particles towards the observation. Because of this correction step, the sampled filtering trajectories between two observation times do not correspond to trajectories of the dynamical model. This implies that from such a particle filter, the standard smoothing based on existing trajectories will not be able to reduce the temporal discontinuities observed on Figure \ref{fig:Vorti_filtering}. This can be observed on Figure \ref{fig:Vorti_standard_smoothing}, where smoothed vorticity maps are computed as $\sum_{i=1}^{N}w_{400}^{(i)} \boldsymbol \xi_{t}^{(i)}$ for $t=400$, and as $\sum_{i=1}^{N}w_{500}^{(i)} \boldsymbol \xi_{t}^{(i)}$ for all $t \in ]400,500]$. The discontinuity at time $t=500$ is still present.

~\\

~\\

\begin{figure}[H]
\centering
\begin{tabular}{cc}
\begin{tabular}{ccc}
   \includegraphics[width=3.5cm,height=3.5cm]{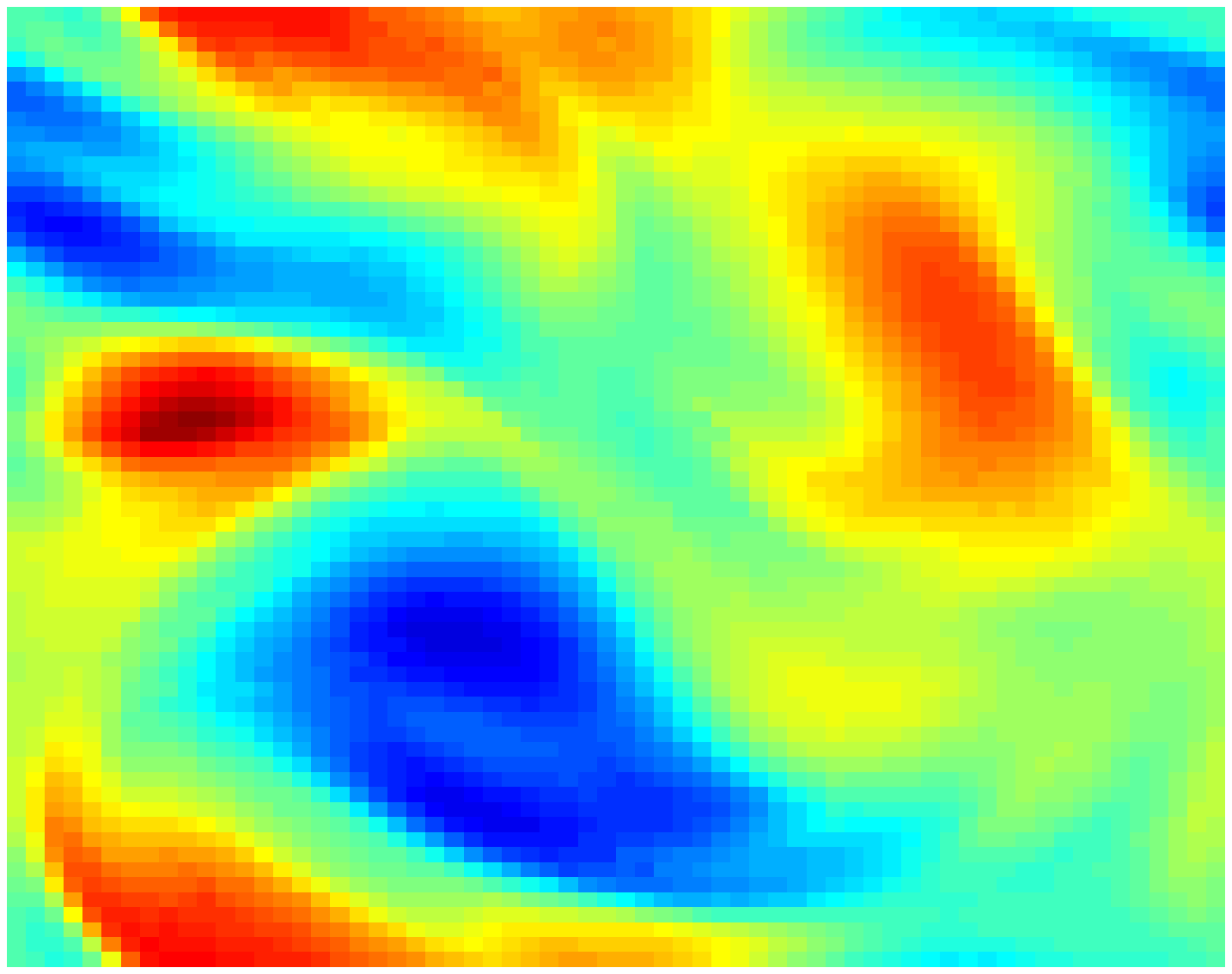} &
   \includegraphics[width=3.5cm,height=3.5cm]{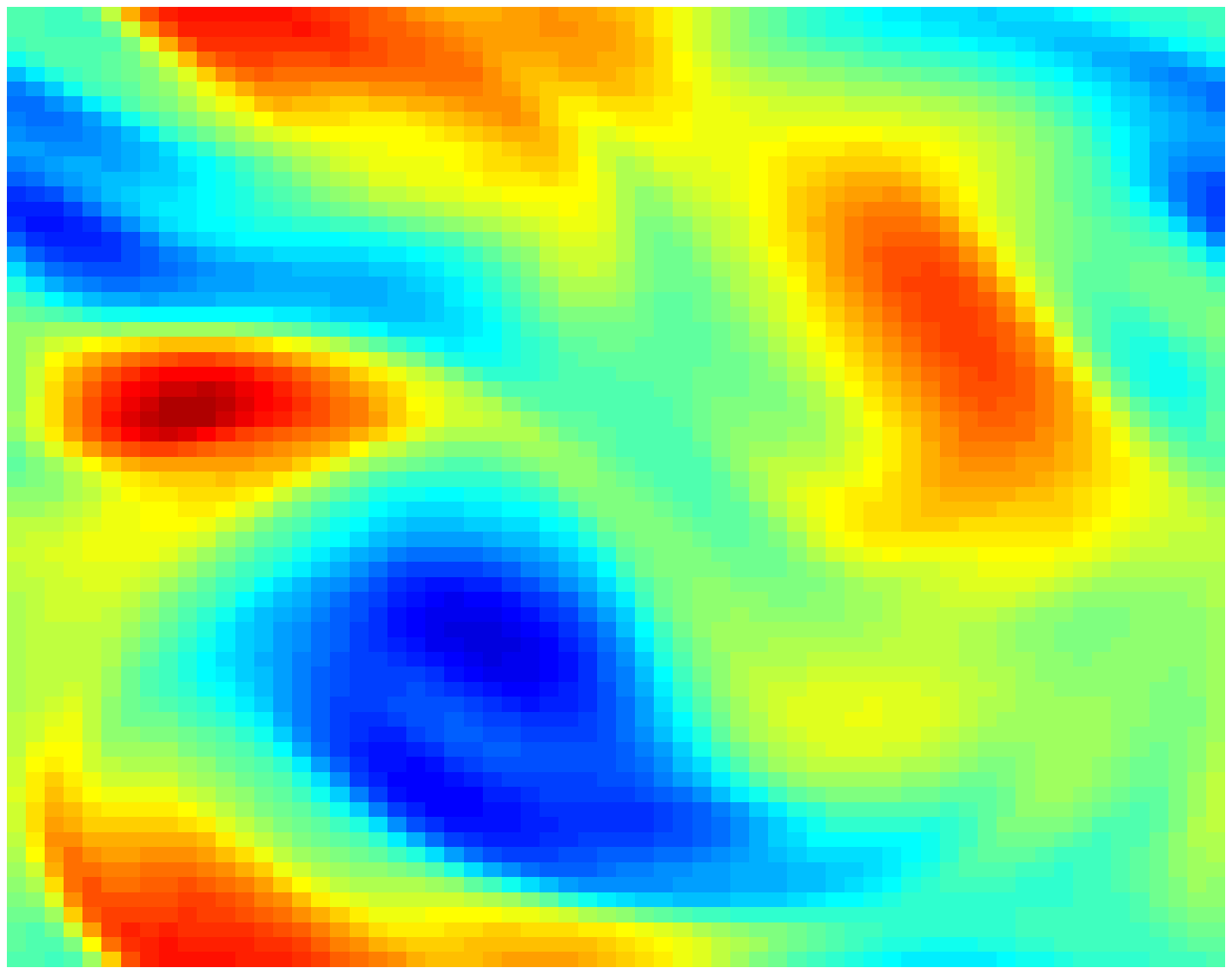} &
   \includegraphics[width=3.5cm,height=3.5cm]{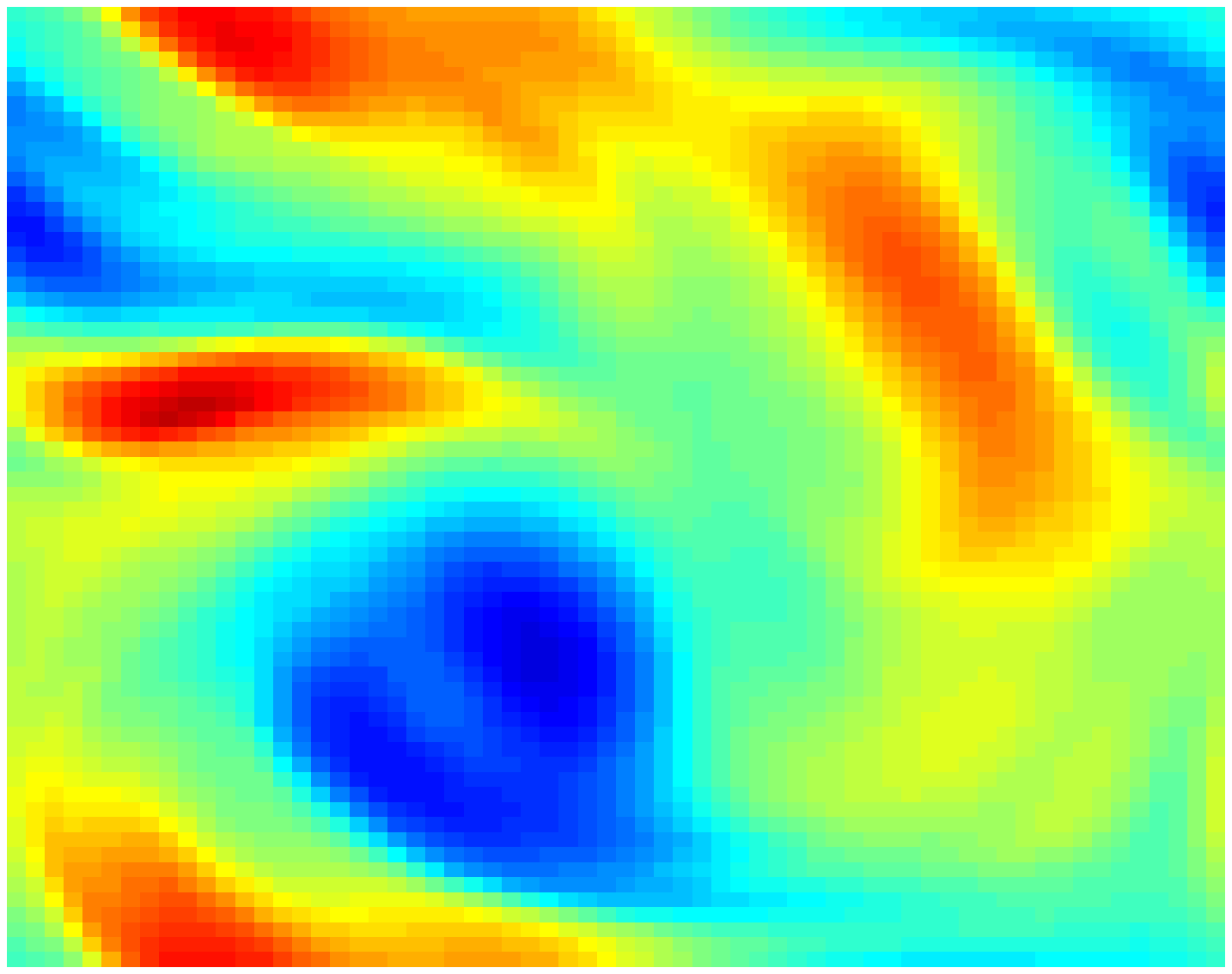} \\
     \footnotesize{$t=400$} & 
     \footnotesize{$t=420$} & 
     \footnotesize{$t=450$} \\
   \includegraphics[width=3.5cm,height=3.5cm]{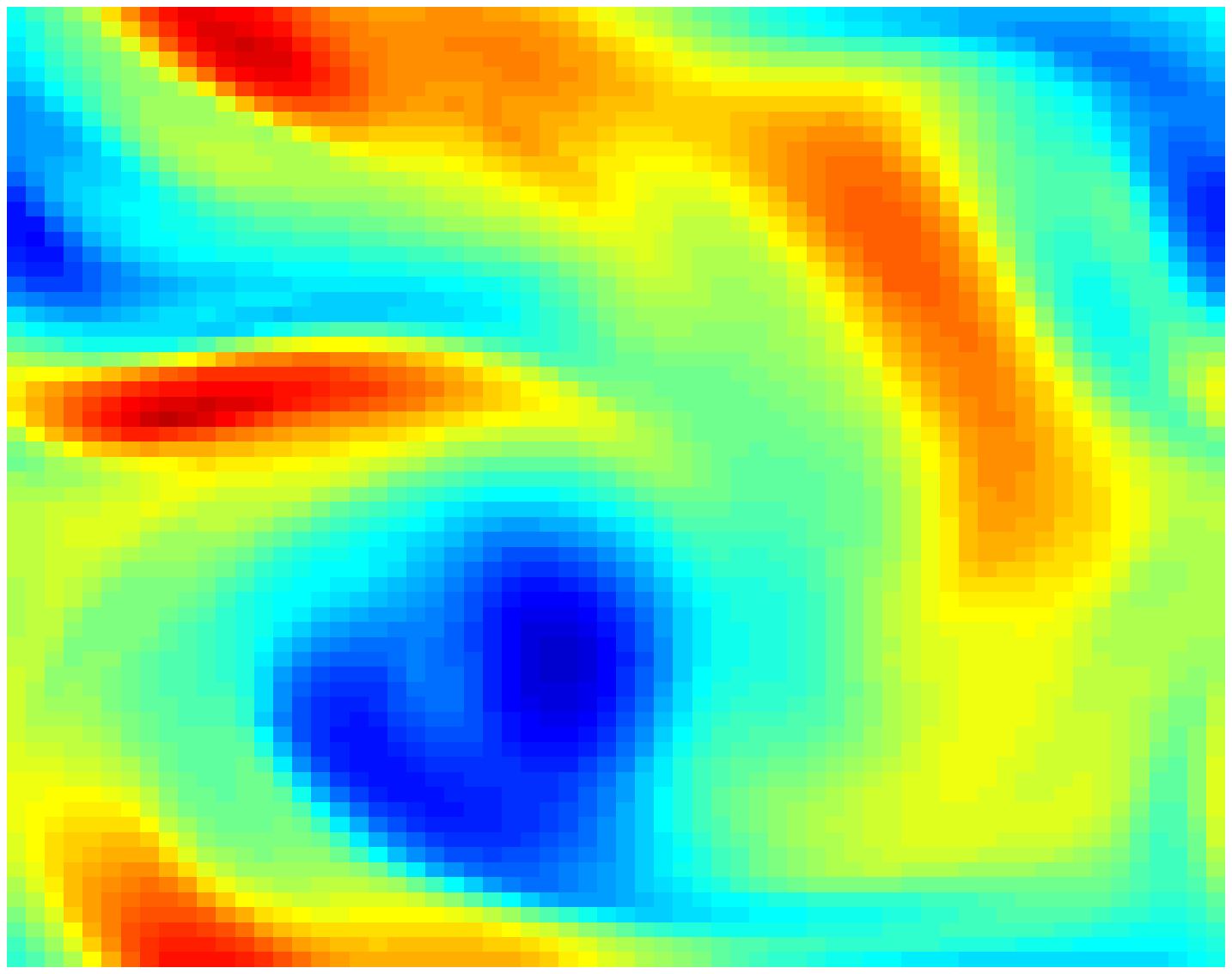} &
   \includegraphics[width=3.5cm,height=3.5cm]{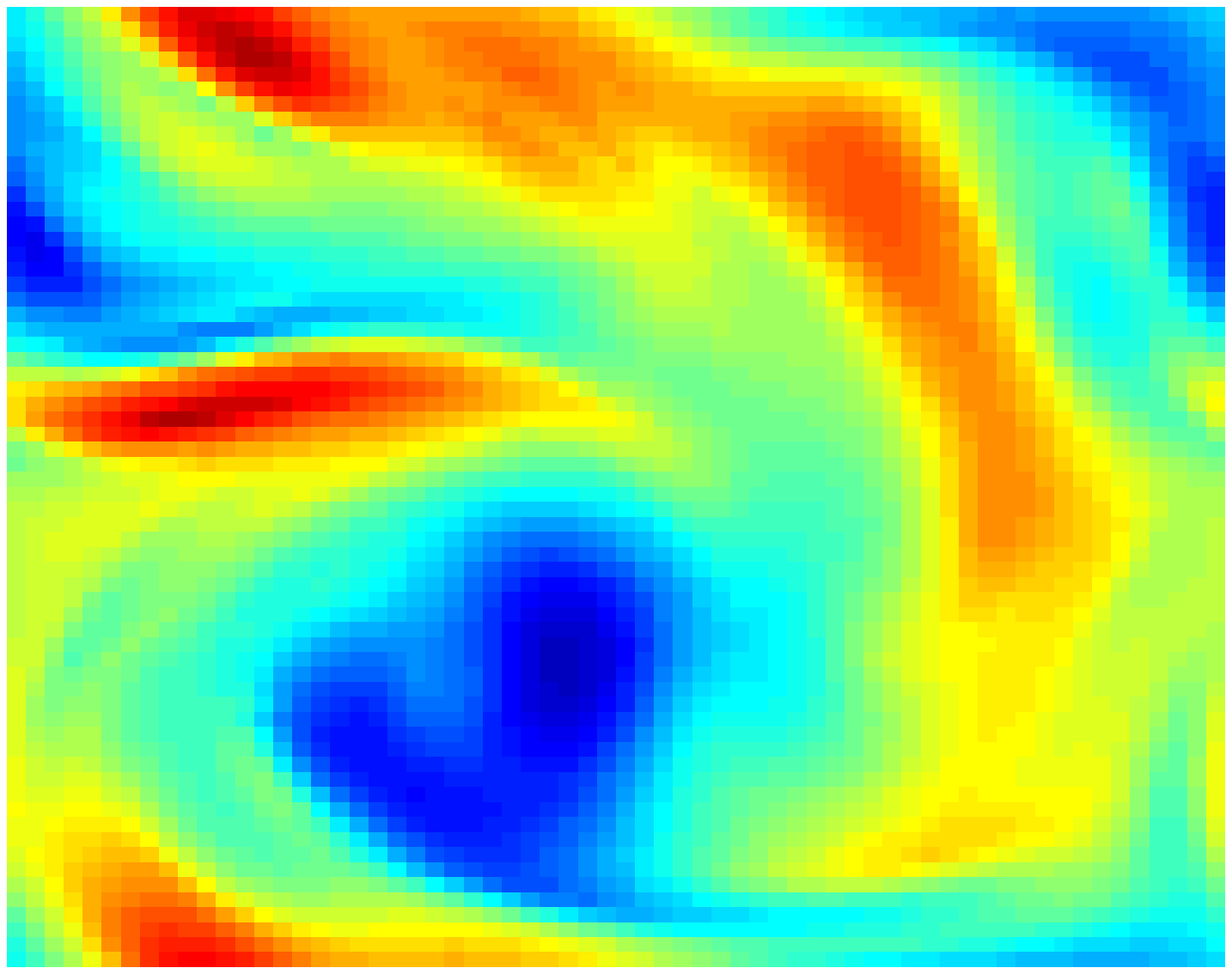} &
   \includegraphics[width=3.5cm,height=3.5cm]{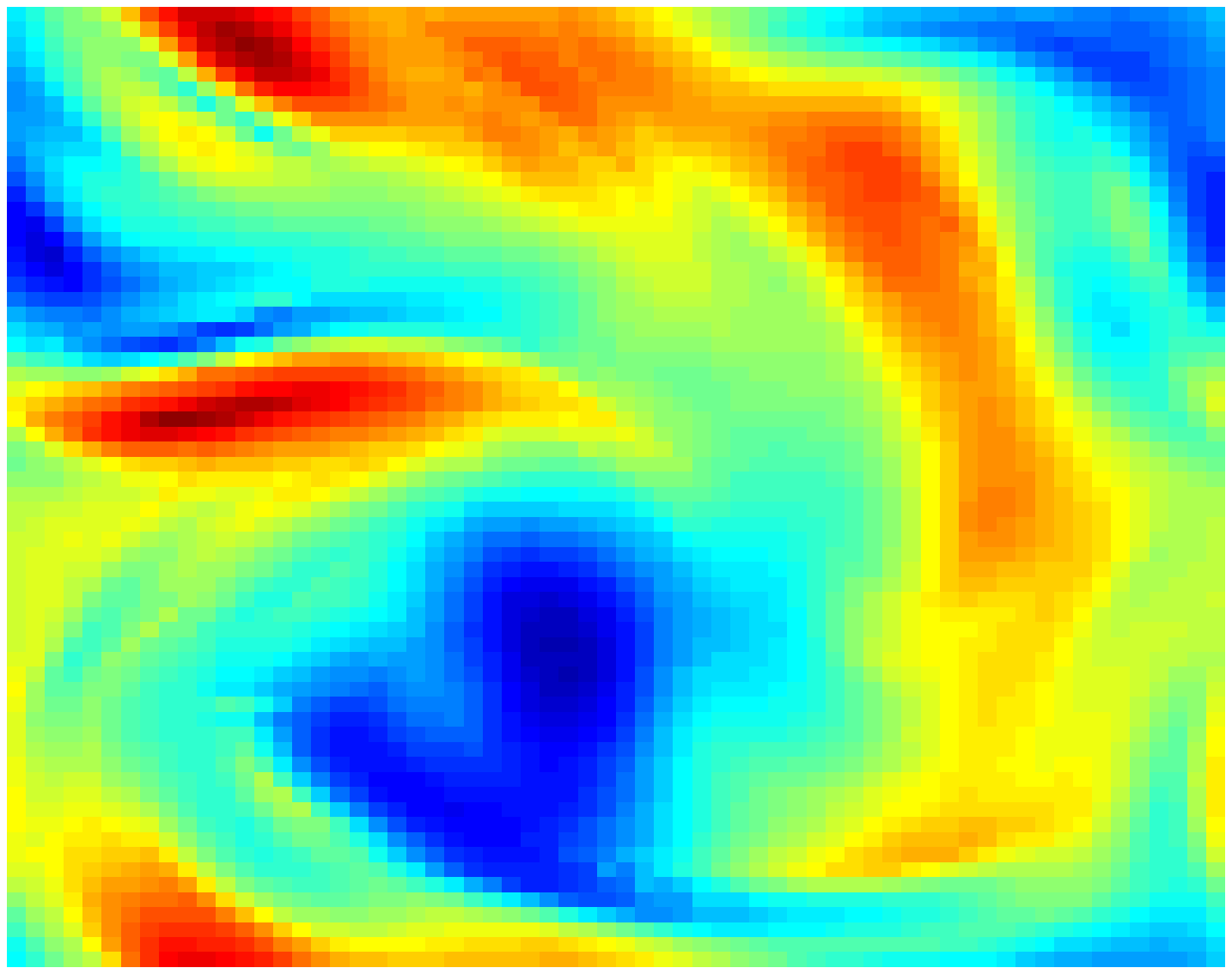} \\
 \footnotesize{$t=470$} & 
     \footnotesize{$t=490$} & 
    \footnotesize{$t=500$} \\
    \end{tabular}
    &
     \begin{tabular}[]{c} 
\includegraphics[width=0.5cm,height=3.5cm]{Scale_vorti.eps} 
\end{tabular}
    \end{tabular}
\caption{Smoothing result with the proposed method. Estimated  mean vorticity maps for different times $t$ between observation times $t=400$ and $t=500$.}
\label{fig:Vorti_smoothing}
\end{figure}

\noindent The result obtained with the proposed method is plotted on Figure \ref{fig:Vorti_smoothing}. Estimated mean vorticity maps are computed as $\sum_{i=1}^N w_{500}^{(i)} \sum_{j=1}^M  \alpha(\tilde{\boldsymbol \xi}^{(i)(j)}) \tilde{\boldsymbol \xi}_t^{(i)(j)}$ for all $t \in [400,500]$. Spatio-temporal vorticity trajectories are gradually modified until observation time $t=500$,   preserving the fluid flow properties. As a matter of fact, since the proposed method samples new trajectories from the law of the physical process (\ref{Navier_Stokes_stochastic}),  the smoothed vorticity trajectories  are by construction consistent with the \textit{a priori} dynamical model. In order to sample the smoothed trajectories, the method relies on the model and on filtering marginals at observation times, but not on filtering trajectories at hidden times. It is then able to smooth the discontinuities inherent to the particle filtering technique we have used, contrary to the standard  smoothing presented on Figure \ref{fig:Vorti_standard_smoothing}.

~\\

\conclusions[Conclusion and discussion]
\label{Discussion}

This paper has introduced a smoothing algorithm based on a conditional simulation technique of diffusions.  The proposed smoothing is formulated as fixed-lag, in the sense that it is performed sequentially each time a new observation appears, in order to correct the state at hidden times up to the previous observation. Note that a decomposition similar to equations (\ref{eq_1})  to (\ref{Conditional_smoothing_particles}) can be written from an integration up to a previous time $t_{k-h}$, with $h>1$. This implies that the smoother can be formulated with a larger fixed-lag, in order to correct the state backward not only up to the previous observation, but up to further measurement times. Yet, due to the successive resampling steps that have been performed in the filtering steps before time $t_k$, there are in practice only a few distinct filtering trajectories at times $t_{k-h}$ if $h$ is large.  Consequently, the estimation of the joint law in (\ref{eq_2}) will not be reliable anymore for a too large value of $h$. 

We have shown the practical applicability of the method to a high-dimensional problem. Nevertheless, the algorithm remains  costly since a second Monte Carlo step is added to the Monte Carlo nature of particle filter algorithms. Yet, from an algorithmic point of view, the sequential nature of the proposed technique allows the smoothing to be implemented with a similar structure as  filtering methods (sequential sampling and weighting of model trajectories). It is then easy to couple this smoothing to an operational filtering system and benefit from parallelization strategies for instance.

\bibliographystyle{copernicus}
\bibliography{Simu_cond}

\end{document}